\documentclass[a4paper,11pt]{article}
\usepackage{jheppub}

\usepackage[utf8]{inputenc}
\usepackage{amsmath}
\usepackage{amssymb}
\usepackage{mathrsfs}
\usepackage{graphicx}
\usepackage{bbold}
\usepackage{comment}
\usepackage[dvipsnames]{xcolor}
\usepackage{braket}
\usepackage{bm,bbm}
\usepackage{physics}
\usepackage{tikz}

\newcommand\beq{ \begin{eqnarray} }
\newcommand\eeq{ \end{eqnarray} }
\newcommand{\SU}{\mathrm{SU}}

\preprint{YITP-23-98, RIKEN-iTHEMS-Report-23}

\arxivnumber{2307.16655}

\title{\boldmath Calculating composite-particle spectra in Hamiltonian formalism and demonstration in 2-flavor QED$_{1+1\text{d}}$}

\author[a,b]{Etsuko Itou,}
\author[a,b]{Akira Matsumoto}
\author[a]{and Yuya Tanizaki}

\affiliation[a]{
Yukawa Institute for Theoretical Physics, Kyoto University,
Sakyo-ku, Kyoto 606-8502, Japan}

\affiliation[b]{
Interdisciplinary Theoretical and Mathematical Sciences Program (iTHEMS), RIKEN,\\
Wako 351-0198, Japan}

\emailAdd{itou(at)yukawa.kyoto-u.ac.jp}
\emailAdd{akira.matsumoto(at)yukawa.kyoto-u.ac.jp}
\emailAdd{yuya.tanizaki(at)yukawa.kyoto-u.ac.jp}

\abstract{
We consider three distinct methods to compute the mass spectrum of gauge theories in the Hamiltonian formalism: (1)~correlation-function scheme, (2)~one-point-function scheme, and (3)~dispersion-relation scheme.
The first one examines spatial correlation functions as we do in the conventional Euclidean Monte Carlo simulations.
The second one uses the boundary effect to efficiently compute the mass spectrum. 
The third one constructs the excited states and fits their energy using the dispersion relation with selecting quantum numbers. 
Each method has its pros and cons, and we clarify such properties in their applications to the mass spectrum for the 2-flavor massive Schwinger model at $m/g=0.1$ and $\theta=0$ using the density-matrix renormalization group (DMRG). 
We note that the multi-flavor Schwinger model at small mass $m$ is a strongly coupled field theory even after the bosonizations, and thus it deserves to perform the first-principles numerical calculations. 
All these methods mostly agree and identify the stable particles, pions $\pi_a$ ($J^{PG}=1^{-+}$), sigma meson $\sigma$ ($J^{PG}=0^{++}$), and eta meson $\eta$ ($J^{PG}=0^{--}$). 
In particular, we find that the mass of $\sigma$ meson is lighter than twice the pion mass, and thus $\sigma$ is stable against the decay process, $\sigma \to \pi\pi$. 
This is consistent with the analytic prediction using the WKB approximation, and, remarkably, our numerical results are so close to the WKB-based formula between the pion and sigma-meson masses, $M_\sigma/M_\pi=\sqrt{3}$.}

\begin{document}
\maketitle
\flushbottom

\section{Introduction}
\label{sec:intro}

In recent years, numerical simulations of quantum field theories (QFTs) in the Hamiltonian formalism have attracted a lot of attention motivated by the rapid progress of quantum computing technology and also the developments of tensor network techniques.
These methods rely on different disciplines from that of Monte Carlo simulations for the conventional lattice gauge theories, and thus they are expected to give complementary frameworks. 
One of the remarkable features is that these methods do not rely on importance sampling, and thus we may be able to circumvent the issue of sign problems.

With this motivation in mind, we investigate the methods to calculate the mass spectrum of gauge theories in Hamiltonian formalism. 
When studying strongly coupled QFTs, we often encounter situations where the fundamental degrees of freedom defining the theory do not appear in the low-energy spectrum. 
Quantum chromodynamics (QCD) is a notable, successful example of such phenomena: Quarks and gluons are confined inside the color-singlet hadrons, and it explains the physics of strong interaction. 
The Monte Carlo simulations nicely predict the hadron spectrum \cite{FlavourLatticeAveragingGroupFLAG:2021npn}
and also the physics at finite temperature \cite{Borsanyi:2013bia,HotQCD:2014kol} in the sign-problem-free regions. 
Of course, we are currently very far away to reproduce such tremendous achievements of Monte Carlo simulations, and thus it is important to develop the counterparts of those calculational techniques in Hamiltonian formalism. 

In this work, we consider three independent methods to compute the mass spectrum of the 2-flavor massive Schwinger model using the density-matrix renormalization group (DMRG):
\begin{itemize}
    \item correlation-function scheme
    \item one-point-function scheme 
    \item dispersion-relation scheme 
\end{itemize}
The first one examines spatial correlation functions as we do in the conventional Euclidean Monte Carlo simulations.
The second one uses the boundary effect for efficiently computing the mass spectrum, which is partly motivated by the applications of the Friedel oscillations~\cite{PhysRevB.54.13495, SHIBATA19971024}. 
The third one constructs the excited states and fits their energy using the dispersion relation with selecting quantum numbers (see, e.g., Refs.~\cite{Pirvu_2012, Haegeman_2012, Haegeman_2013} for similar analysis in spin systems). 
Each method has its pros and cons, and especially the last one is specific to the Hamiltonian formalism. 
Our purpose is to clarify their properties in the concrete studies of the 2-flavor massive Schwinger model. 

The Schwinger model is a $1+1$d quantum electrodynamics (QED$_{1+1\text{d}}$)~\cite{Schwinger:1962tp}, and it is a strongly-coupled theory like $4$d QCD: The fundamental fermions are confined because of the linear Coulomb potential, and the low-lying states are composite states like mesons. 
Despite its strong-coupling nature, one can calculate many nontrivial aspects using analytical methods~\cite{Lowenstein:1971fc, Casher:1974vf, Coleman:1975pw, Coleman:1976uz, Manton:1985jm, Hetrick:1988yg, Jayewardena:1988td, Sachs:1991en, Adam:1993fc, Adam:1994by, Hetrick:1995wq, Narayanan:2012du, Narayanan:2012qf, Lohmayer:2013eka, Tanizaki:2016xcu}, 
and this theory has been used as a benchmark to test new computational methods in previous studies (see, e.g., Refs.~\cite{Banuls:2013jaa, Banuls:2015sta, Banuls:2016lkq, Buyens:2016ecr, Buyens:2016hhu, Banuls:2016gid, Funcke:2019zna, Chakraborty:2020uhf, Honda:2021aum, Honda:2021ovk, Honda:2022edn, Tomiya:2022chr, Funcke:2023lli, Dempsey:2023gib,  Kharzeev:2020kgc, deJong:2021wsd, Nguyen:2021hyk, Nagano:2023uaq} for numerical studies of Schwinger model with tensor networks and/or quantum simulations). 
The mass spectrum of this model was studied numerically by the Monte Carlo method as well \cite{Fukaya:2003ph}, including the region with nonzero $\theta$ angles with the reweighting technique. 
As another Monte Carlo based studies, the dual variable formulations are developed, which successfully eliminates the sign problem of $1+1$d U(1) gauge theories~\cite{Gattringer:2015nea, Gattringer:2015baa, Gattringer:2018dlw}. 
There is also a numerical approach with the light-cone quantization in the Hamiltonian formalism~\cite{Harada:1993va} by using the so-called Tamm-Dancoff approximation.

The low-energy mass spectrum of the 2-flavor Schwinger model with a theta term has been studied analytically
by using the bosonization technique \cite{Coleman:1976uz, Hetrick:1995wq}. 
We note, however, that the low-energy effective theory is strongly coupled if $0<m/g\ll1$ even after bosonization, and the details of the prediction rely on some approximations that are not fully justifiable. 
It is physically nontrivial if those analytic predictions are reproduced by the first-principles numerical computations when we go into the details beyond qualitative aspects. 

We performed the DMRG computations using the C++ library of ITensor \cite{itensor} to obtain the mass spectrum at $\theta=0$ with the above three methods. 
We find that all three methods mostly agree and identify the stable particles, pions $\pi_a$ ($J^{PG}=1^{-+}$), sigma meson $\sigma$ ($J^{PG}=0^{++}$), and eta meson $\eta$ ($J^{PG}=0^{--}$), where $J$ denotes the isospin quantum number, $P$ and $G$ is the parity and $G$-parity, respectively. 
In particular, we observe that the mass of $\sigma$ meson is lighter than twice the pion mass, and thus $\sigma$ is stable against the decay process, $\sigma \to \pi\pi$. 
This implies that $\sigma$ is a stable particle, not a $\pi\pi$ resonance, and this is a notable difference compared with $4$d QCD.
This is consistent with the analytic prediction based on the WKB approximation of the Abelian bosonized description. 
The WKB-based formula predicts the $\pi$ and $\sigma$ mass ratio is given by $M_\sigma/M_\pi=\sqrt{3}$, 
and our three distinct computations give roughly consistent results:\footnote{
The errors of these values correspond to the fitting error in each scheme.
These values should be further affected by systematic errors potentially coming from, for instance, the finite lattice spacing, the finite-volume effect,
the effect of the open boundary, and the cutoff on the bond dimension.}
\begin{equation}
    \begin{array}{c|c|c|c}
        & \mbox{correlation-function} & \mbox{one-point-function} & \mbox{dispersion-relation}\\ \hline
        M_\sigma/M_\pi & 1.68(2) & 1.821(6) &  1.75(1)   
\end{array}
\end{equation}
Let us emphasize that this poses an interesting theoretical question on why the semiclassical prediction works so well even outside its valid regime.
In this paper, we concentrate on the simulation at $\theta=0$ to confirm
the validity of our method. In fact, it should be straightforward to extend our work
to the $\theta\neq0$ region.

This paper is organized as follows.
In Section~\ref{sec:review}, we review the continuum 2-flavor Schwinger model and the bosonization analysis focusing on the mass spectrum.
In Section~\ref{sec:formulation}, we introduce the lattice formulation of the Hamiltonian and define some observables.
In Section~\ref{sec:dmrg}, we briefly explain our simulation method, the DMRG algorithm and show our setup of the simulation.
In Section~\ref{sec:result}, we present our simulation results for the three methods.
Section~\ref{sec:summary} is devoted to the conclusion and discussion.
Appendix~\ref{sec:Spin_op} shows the explicit form of the Hamiltonian and the observables in the spin representation for DMRG.
In Appendix~\ref{sec:C_test}, we test the validity of the charge conjugation operator on the lattice in the 1-flavor Schwinger model.
In Appendix~\ref{sec:order_of_index}, we discuss the assignment of the flavor index for constructing MPS.
In Appendix~\ref{sec:CF_nf1}, we investigate how the truncation of the bond dimension affects the correlation function in the massless 1-flavor Schwinger model.

\section{Review of the 2-flavor Schwinger model}
\label{sec:review}

In this work, we study the 2-flavor Schwinger model, which is $(1+1)$-dimensional quantum
electrodynamics (QED$_{1+1\text{d}}$) with $N_{f}=2$ species of Dirac fermion.
The Lagrangian density with the Minkowski metric
$\eta_{\mu\nu}=\mathrm{diag}(1,-1)$ is given by 
\begin{equation}
\mathcal{L}=-\frac{1}{4g^2}F_{\mu\nu}F^{\mu\nu}+\frac{\theta}{4\pi}\epsilon_{\mu\nu}F^{\mu\nu}+\sum_{f=1}^{N_{f}}\left[i\bar{\psi}_{f}\gamma^{\mu}\left(\partial_{\mu}+i A_{\mu}\right)\psi_{f}-m\bar{\psi}_{f}\psi_{f}\right],
\label{eq:Lagrangian}
\end{equation}
where $F_{\mu\nu}=\partial_{\mu}A_{\nu}-\partial_{\nu}A_{\mu}$ is
the field strength, $g$ is the gauge coupling, and $\theta$ is the vacuum angle describing the background electric flux. 
The index $f$ labels the flavor.
We set the masses of the two fermions equal to $m$.

\subsection{Global symmetry and composite operators}
\label{subsec:def-composite}

In the chiral limit ($m=0$), the 2-flavor Schwinger model has the chiral symmetry and the $G$-parity symmetry,
\begin{equation}
    \frac{\SU(2)_L\times \SU(2)_R}{\mathbb{Z}_2}\times (\mathbb{Z}_2)_G\quad (m=0),
\end{equation}
and the chiral symmetry has an 't~Hooft anomaly. 
The $G$-parity operation is the combination of the charge-conjugation with the $\pi$ rotation of the $\SU(2)_V$, which will be discussed later. 
We note that the continuous chiral symmetry cannot be spontaneously broken due to the Coleman-Mermin-Wagner theorem, and the anomaly matching condition is satisfied by the $\SU(2)$ level-$1$ Wess-Zumino-Witten ($\SU(2)_1$ WZW) conformal field theory. 
The $\SU(2)_1$ WZW model is equivalent to the self-dual compact boson, and one can explicitly derive it from the massless 2-flavor Schwinger model with the Abelian bosonization~\cite{Coleman:1976uz}. 

The massive 2-flavor Schwinger model \eqref{eq:Lagrangian} no longer has the chiral symmetry, but it maintains the following symmetry, 
\begin{equation}
    \left\{
    \begin{array}{cc}
    [\SU(2)_V/\mathbb{Z}_2]\times (\mathbb{Z}_2)_G     &  \quad (\theta=0 \bmod 2\pi),\\{}
    [\SU(2)_V/\mathbb{Z}_2]\times (\mathbb{Z}_2)_{G+L} & \quad (\theta=\pi \bmod 2\pi),\\
    \SU(2)_V/\mathbb{Z}_2     & \quad (\text{else}), 
    \end{array} 
    \right.
\end{equation}
and we call $\SU(2)_V/\mathbb{Z}_2$ as the isospin symmetry. $(\mathbb{Z}_2)_{G+L}$ is the diagonal subgroup between the center of $\SU(2)_L$ and the $G$-parity $(\mathbb{Z}_2)_G$. 
The $\mathbb{Z}_2$ quotient of $\SU(2)/\mathbb{Z}_2$ means that the local operators always have the integer isospin quantum numbers since the gauge invariance requires that the local operator must consist of the same number of $\psi$ and $\bar{\psi}$.
We define the isospin operators $J_a$ as conserved charges under this symmetry by
\begin{equation}
J_a = \frac{1}{2}\int dx\, \bar{\psi} \gamma^0 \tau_a \psi,
\label{eq:J_cont}
\end{equation}
where $\tau_a$ represents Pauli matrices of the isospin space with $a\in\{x,y,z\}$.

When $\theta$ takes some special values, e.g.  $\theta=0$, the theory enjoys the charge conjugation,
\begin{equation}
    C: A\to -A, \quad \psi\leftrightarrow \mathsf{C}\overline{\psi}^t, 
\end{equation}
with a suitable element of the Clifford algebra $\mathsf{C}$. 
We note that this operation flips the sign of the $\theta$ angle, and thus this symmetry does not exist for generic values of $\theta$. 
For general numbers of flavors, $C$ acts on $\SU(N_f)/\mathbb{Z}_{N_f}$ as an outer automorphism, i.e. the symmetry group becomes $[\SU(N_f)/\mathbb{Z}_{N_f}]\rtimes (\mathbb{Z}_2)_C$. 
When $N_f=2$, however, $\SU(2)$ does not have nontrivial outer automorphisms, and indeed $C$ just gives the $\pi$ rotation in the isospin space.  
Thus, it is convenient to introduce the $G$-parity \cite{Gparity},
\begin{equation}
G=Ce^{i\pi J_{y}},\label{eq:G_parity}
\end{equation}
so that it commutes with the isospin operation and gives a well-defined eigenvalue $\pm1$.
Moreover, the $G$-parity acts trivially on the $\SU(2)_1$ WZW theory. Thus, if we find a particle with $G=-1$, we can immediately tell it remains massive in the chiral limit.

In this paper, we mainly focus on the following composite operators to discuss the meson spectrum:
\begin{alignat}{3}
    &\pi_a&&=-i \bar{\psi}\gamma^5 \tau_a \psi \quad  && (J^{PG}=1^{-+}),
    \label{eq:pi_meson} 
    \\
    &\sigma&&=\bar{\psi}\psi && (J^{PG}=0^{++}),
    \label{eq:sigma_meson}
    \\
    &\eta&&=-i \bar{\psi}\gamma^5\psi && (J^{PG}=0^{--}). 
    \label{eq:eta_meson}
\end{alignat}
We call them pion, sigma, and eta operators, respectively, obviously motivated by the meson spectrum of $4$d QCD. We will often denote $\pi_3=\pi$ for simplicity. 
Here, we have specified their quantum numbers $J^{PG}$, where $J$ denotes the isospin, and $P$ and $G$ denote the parity and the $G$-parity at $\theta=0$, respectively. 
The $(1+1)$d QED is strongly coupled, and it turns out that the light particles correspond to these operators, and this feature is reminiscent of $4$d QCD.  
Here, we would like to note that $\eta$ has $G=-1$, and thus it remains massive in the chiral limit, which is analogous to the $U(1)_A$ problem~\cite{Coleman:1975pw, Coleman:1976uz, Frohlich:1976mt}. 
Other mesons, $\pi$ and $\sigma$ have $G=+1$ and actually become massless in the chiral limit. 
The massless $\sigma$ particle is an outcome of the absence of chiral symmetry breaking, unlike the $4$d QCD case.

\subsection{Phase structure}
\label{subsec:phase_structure}

With $m\not =0$, the system is gapped and has the unique ground state at generic values of $\theta$. 
In $(1+1)$d, there is no stable topologically-ordered state, and the unique gapped ground states are then classified as the symmetry-protected topological (SPT) states~\cite{PhysRevB.83.035107, Kapustin:2014gma, Kapustin:2014tfa}. 
This perspective provides a very powerful tool to understand the phase structure of the 2-flavor Schwinger model. 

Let us recall that the massive Schwinger model always has the isospin symmetry, $\SU(2)_V/\mathbb{Z}_2$. 
We can then calculate the partition function under the presence of the background gauge fields for the isospin symmetry. 
Compared with the $\SU(2)$ gauge field, the $\SU(2)/\mathbb{Z}_2$ gauge field has milder cocycle conditions, which is controlled by the $\mathbb{Z}_2$ $2$-form gauge field $w_2$ in addition to the familiar $1$-form gauge field. As a result, at generic values of $\theta$, the partition function with the background gauge field is described by the low-energy effective topological action,  
\begin{equation}
    \mathcal{Z}_{\theta}\simeq \exp(i \pi k \int w_2), 
\end{equation}
with some $k\sim k+2$. We note that $k$ is a discrete label that distinguishes the SPT states protected by $\SU(2)_V/\mathbb{Z}_2$, and thus it cannot be changed under the continuous change of coupling constants unless quantum phase transitions happen. 

We can prove that two partition functions at $\theta$ and $\theta+2\pi$ are different in the presence of the background gauge fields. The anomalous relation can be summarized as~\cite{Misumi:2019dwq} 
\begin{equation}
    \mathcal{Z}_{\theta+2\pi}=\exp\left(i \pi \int w_2\right) \mathcal{Z}_{\theta}. 
\end{equation}
The label $k$ is changed as $k\mapsto k+1$ as we change the $\theta$ angle from $\theta$ to $\theta+2\pi$, and there must be a phase transitions separating the $k=0,1$ ground states. 
It is somewhat customary to assign the $k=0$ state for $-\pi<\theta<\pi$ and the $k=1$ state for $\pi<\theta<3\pi$, while, precisely, this assignment depends on the UV-regularization scheme. 
We note that the whole story here is quite parallel to that of the anti-ferromagnetic Heisenberg chain or the $(1+1)$d $\mathbb{C}P^1$ sigma model~\cite{Haldane:1983ru, Affleck:1986pq, Haldane:1988zz, Affleck:1987vf, Komargodski:2017dmc, Komargodski:2017smk, Lajko:2017wif, Tanizaki:2018xto} (except for the properties at $\theta=\pi$~\cite{Coleman:1976uz, Dempsey:2023gib}). 

The distinction between the states at $\theta=0$ and $\theta=2\pi$ becomes more vivid when we take the open boundary condition. 
Turning on non-zero $\theta$ corresponds to introducing a background electric field with a constant magnitude $\theta/2\pi$.
When we increase $\theta$ beyond $\pi$, the background field becomes larger than $1/2$.
Then the Dirac fermions with charges $\pm1$ are excited at the boundaries to cancel the background field as much as possible.
As a consequence, these boundary states have isospin $1/2$, which is the projective representation of $\SU(2)_V/\mathbb{Z}_2$. 
This is nothing but the signature of the nontrivial SPT state. 
If the system size is large enough, the interaction between the boundary states is exponentially suppressed, 
so the two independent degrees of freedom with isospin $1/2$ yield $2\times 2$ degeneracy of the ground state.
We will use this boundary excitation as a source of iso-triplet particles to determine
the mass of the pion from the one-point function in Section \ref{subsec:result_1pt}.
Note that if we increase $\theta$ further beyond $3\pi$, the boundary excitations become bound states of Dirac fermions with isospin 1, 
which can be completely screened by the gauge-invariant particles inside the bulk, and  the ground state would be unique again with the open boundary condition.

\subsection{Mass spectrum}
\label{subsec:mass_spectrum}

In this subsection, we are going to give a relatively detailed review on analytic predictions about the mass spectrum. 
There is already a huge effort for the analytic studies of the multi-flavor Schwinger model, and thus the reader may wonder how one could obtain something new with numerical studies. 
We would like to clarify what kinds of approximations were used in the previous studies and give the justification of this work on physical aspects. 

There are two exactly-solvable limits of the multi-flavor Schwinger model: 
\begin{itemize}
    \item In the heavy fermion limit $m\to \infty$, the model becomes the pure $U(1)$ gauge theory. 
    \item In the chiral limit $m=0$, the model becomes the $SU(N_f)_1$ WZW conformal field theory and one massive free boson, and they are completely decoupled. 
\end{itemize}
It is then natural to consider perturbations from these limits in order to investigate the cases of general mass $m$. 
When $m\gg g$, one can perform systematic perturbations to study the spectrum. 
In the opposite case $0< m\ll g$, however, the systematic perturbation works only for the $1$-flavor case,
and further approximations are necessary for $N_f\ge 2$.

By applying the Abelian bosonization for $N_f=2$, the fermions are mapped to the $2\pi$-periodic scalar fields $\phi_1, \phi_2$. 
The Lagrangian~\eqref{eq:Lagrangian} is then completely equivalent to 
\begin{align}
    \mathcal{L}&=\frac{1}{2g^2}F_{01}^2+\frac{1}{2\pi}(\phi_1+\phi_2+\theta)F_{01}
    \notag\\
    &+\frac{1}{8\pi}\left( (\partial \phi_1)^2+(\partial \phi_2)^2\right)+ C m \rho N_\rho[\cos(\phi_1)+\cos(\phi_2)], 
\end{align}
where $C=e^\gamma/(2\pi)$ is a numerical constant, and $N_\rho$ denotes the normal ordering for the contraction with a free field propagator of mass $\rho$ \cite{Coleman:1974bu}.\footnote{All the UV divergences from loop diagrams are removed by this prescription, and the theory is independent of the choice of $\rho$. 
For the free theory with mass $\rho$, $N_{\rho}[\bullet]$ becomes the ordinary normal ordering.} 
We can rigorously integrate out the gauge fields as the Lagrangian is quadratic in terms of $F_{01}$. 
Changing the basis of bosons as $\phi_{1,2}=\sqrt{2\pi}\eta-\frac{\theta}{2}\pm\varphi$,
the effective Lagrangian becomes 
\begin{align}
    \mathcal{L}_{\mathrm{eff}}[\eta,\varphi]=\frac{1}{2}\left[(\partial \eta)^2-\mu^2 \eta^2\right]+\frac{1}{4\pi}(\partial \varphi)^2+2C m \rho N_\rho\left[\cos\left(\sqrt{2\pi}\eta-\frac{\theta}{2}\right)\cos(\varphi)\right], 
    \label{eq:H_bosonization}
\end{align}
where $\mu^2=2g^2/\pi$. 
We have the $\mathbb{Z}_2$ symmetry acting only on $\eta$ when $\theta=0$, and this is the $G$-parity. 
When $m=0$, the massive $\eta$ and the massless $\varphi$ decouple, as advocated above. 

Let us now turn on the small mass, $0<m\ll g$. 
The $\eta$ particle has the mass $\mu+O(m)$, but it is hard to compute the $O(m)$ correction due to the potential infrared divergence in the loop diagrams with the $\varphi$ fields~\cite{Coleman:1976uz}. 
Instead, we integrate out $\eta$ at the tree level to discuss the physics of $\pi$ and $\sigma$ mesons, which gives $\left\langle N_\rho\left[\cos\left(\sqrt{2\pi}\eta-\frac{\theta}{2}\right)\right]\right\rangle = \sqrt{\frac{\mu}{\rho}}\cos \frac{\theta}{2}$ for the free massive $\eta$ with mass $\mu$. 
The effective theory for $\varphi$ becomes the sine-Gordon model,
\begin{equation}
    \mathcal{L}_{\mathrm{SG}}[\varphi]=\frac{1}{4\pi}(\partial \varphi)^2 
    +2Cm \cos\frac{\theta}{2}  (\mu \rho)^{1/2} N_\rho[\cos \varphi]. 
    \label{eq:effLagrangian_phi}
\end{equation}
The isospin $\SU(2)_V/\mathbb{Z}_2$ symmetry is not manifest at all in the Lagrangian, but it secretly exists quantum mechanically. 
In particular, the $z$-component isospin current is given by $J_z^\mu=\frac{1}{2\pi}\varepsilon^{\mu\nu}\partial_\nu\varphi$, and its charge $J_z=\int dx \frac{1}{2\pi}\partial_x \varphi$ counts the winding number of $\varphi\sim \varphi+2\pi$. 

We would like to emphasize that the effective theory~\eqref{eq:effLagrangian_phi} is strongly coupled, and we cannot solve it in the ordinary perturbation for small but nonzero $m$. 
What is actually done in the previous literature is the optimized perturbation; we optimize the renormalization scale $\rho$ so that the coefficient of the $\cos(\varphi)$ potential becomes $O(\rho^2)$, and we get
\begin{equation}
    \rho_{\mathrm{optimzied}}\sim  \left|m\sqrt{\mu}\cos(\theta/2)\right|^{2/3}. 
\end{equation}
This is identified as the mass gap caused by the mass perturbation, and this formula gives the $\theta$-dependence of the lightest meson mass, i.e. $M_\pi$. 

The spectrum of the sine-Gordon model was studied by using WKB approximation~\cite{Dashen:1975hd}. 
Introducing an extra parameter controlling the kinetic term as $\frac{1}{4\pi\beta^2}(\partial \varphi)^2$, the quantum scaling dimension of $\cos\varphi$ becomes $\Delta=\beta^2/2$, so the semiclassical approximation is valid when $\beta^2\to 0$. 
The model has the soliton and antisoliton, and let us denote their mass as $M_{\mathrm{SG}}$. 
Then, Dashen~et~al.~\cite{Dashen:1975hd} predicted the masses of soliton-antisoliton bound states as 
\begin{equation}
    M_{\mathrm{SG}}^{(n)}
    =2M_{\mathrm{SG}}\sin\left(\frac{\pi}{2}\frac{n}{4/\beta^2-1}\right),
    %=2M_{\mathrm{SG}}\sin\left(\frac{n}{16}\,\frac{2\pi\beta^{2}}{1-\beta^{2}/4}\right),
\label{eq:mass_of_SG}
\end{equation}
with $n=1,2,\cdots <  (4/\beta^2 - 1 )$.
Even though it is subtle if the WKB works at the self-dual point $\beta^2=1$, Coleman got an intriguing observation using this semiclassical formula~\cite{Coleman:1976uz}.  

The nontrivial check for its validity is the recovery of the $\SU(2)_V/\mathbb{Z}_2$ symmetry at $\beta^2=1$. 
Substituting $n=1$ with $\beta^2=1$ into \eqref{eq:mass_of_SG}, we get
\begin{equation}
M_{\mathrm{SG}}^{(1)}=2M_{\mathrm{SG}}\sin\frac{\pi}{6}=M_{\mathrm{SG}}.
\end{equation}
This shows that the lightest soliton-antisoliton bound state has the same mass as the soliton or antisoliton itself. 
The soliton and antisoliton have $J_z=\pm 1$, and the soliton-antisoliton bound state has $J_z=0$. 
The $G$-parity does not act on $\varphi$, so all these states have $G=+1$. 
Thus, these three states form the isospin triplet $J^{PG}=1^{-+}$ of mass $M_{\mathrm{SG}}$,
which is identified as the pion in the Schwinger model, and then $M_{\mathrm{SG}}=M_\pi\sim (m\sqrt{\mu}\cos(\theta/2))^{2/3}$. 

The mass of the second soliton-antisoliton bound state is given by 
\begin{equation}
M_{\mathrm{SG}}^{(2)}=2M_{\mathrm{SG}}\sin\frac{\pi}{3}=\sqrt{3}M_{\mathrm{SG}}.\label{eq:sqrt3-SG}
\end{equation}
This state has $J_{z}=0$, and there is no other state with the same mass.
We then identify it as the $\sigma$ meson in the Schwinger model with $J^{PG}=0^{++}$. 
Thus, the semiclassical method predicts that the masses of pion and sigma meson satisfy  
\beq
M_{\sigma} = \sqrt{3}  M_{\pi}.
\label{eq:sqrt3-Schwinger}
\eeq
Importantly, $M_\sigma <2 M_\pi$. 
Unlike the $4$d QCD, $\sigma$ is a stable particle, not a resonance, because the decay $\sigma \to \pi\pi$ is energetically prohibited. 

As we discussed above, the low-energy mass spectra can be estimated by bosonization.
However, it relies on the optimized perturbation and also on the semiclassical method, and these analyses are not necessarily fully justified. 
It is still difficult to compute the exact $m$-dependence or to find other states
with higher energies than $\mu$.
Thus, it is worth studying the mass spectrum of the 2-flavor model by first-principles numerical methods.

\section{Lattice formulation of the 2-flavor Schwinger model}
\label{sec:formulation}

In this section, we explain the Hamiltonian formalism of the 2-flavor Schwinger model
and its lattice regularization as a generalization to $N_f=2$ from previous research
\cite{Chakraborty:2020uhf,Honda:2021aum,Honda:2021ovk,Honda:2022edn}.
We also define various local and global observables used in the analysis.

\subsection{Hamiltonian}

First, we introduce the continuum Hamiltonian of the $N_{f}$-flavor Schwinger model.
By introducing a conjugate momentum $\Pi=\frac{1}{g^2}\partial_{0}A^{1}+\frac{\theta}{2\pi}$,
the Hamiltonian is given by 
\begin{equation}
H=\int dx\,\left\{ \frac{g^2}{2}\left(\Pi-\frac{\theta}{2\pi}\right)^{2}+\sum_{f=1}^{N_{f}}\left[-i\bar{\psi}_{f}\gamma^{1}\left(\partial_{1}+iA_{1}\right)\psi_{f}+m\bar{\psi}_{f}\psi_{f}\right]\right\} .
\label{eq:H_cont}
\end{equation}
In Hamiltonian formalism, the physical Hilbert space is constrained
by the Gauss law condition,
\begin{equation}
\partial_{1}\Pi+\sum_{f=1}^{N_{f}}\psi_{f}^{\dagger}\psi_{f}=0.\label{eq:Gauss_law_continuum}
\end{equation}
The electric field corresponds to $E:=\dot{A}_1=g^2(\Pi-\theta/2\pi)$.
Thus, the theta angle $\theta$ plays the role of the background electric field.
In the periodic boundary condition, the Hamiltonians at $\theta$ and $\theta+2\pi$ are unitary equivalent, $H_{\theta+2\pi}= U^\dagger H_\theta U$ with $U=\exp(-i\int A_1 d x)$, which realizes the $2\pi$ periodicity of $\theta$. 

Next, we consider the lattice regularization of the Hamiltonian.
Here we employ the staggered fermion to define fermions on
the lattice \cite{Kogut:1974ag,Susskind:1976jm}.
The staggered fermions $\chi_{f,n}$ with the lattice spacing $a$ represents the discretization of the two-component Dirac fermions $\psi_{f}(x)$ with the lattice spacing $2a$.
The single-component fermions $\chi_{f,n}$ at the site $n=0,1,\cdots,N-1$ correspond
to each component\footnote{
The labels $u$ and $d$ of $\psi_{f}$ denote the upper and lower spinor component respectively.
They are nothing to do with up and down quark in QCD here.}
of $\psi_{f}(x)$ depending on $n$ as
\begin{equation}
\psi_f(x)=\begin{pmatrix}
    \psi_{u,f}(x)\\
    \psi_{d,f}(x)
\end{pmatrix}\leftrightarrow\frac{1}{\sqrt{2a}}\begin{pmatrix}
    \chi_{f,2[n/2]}\\
    \chi_{f,2[n/2]+1}
\end{pmatrix}. 
\label{eq:chi_to_psi}
\end{equation}
The number of staggered fermions for each flavor is equal to $N$, thus at each site, there are two staggered fermions.
In this work, we set $N$ to be an even number.

The gauge field is encoded to U(1) variables $U_{n}\sim \exp(-iaA^{1}(x))$, defined on the link between the $n$-th and $(n+1)$-th sites,
and the conjugate momentum is replaced by $L_{n}\sim-\Pi(x)$, defined on the $n$-th site. 
The canonical commutation relations are given by
\begin{equation}
\{\chi_{f,n}^{\dagger},\,\chi_{f^{\prime},m}\}=\delta_{ff^{\prime}}\delta_{nm},\label{eq:CC_relation_1}
\end{equation}
\begin{equation}
\{\chi_{f,n},\,\chi_{f^{\prime},m}\}=\{\chi_{f,n}^{\dagger},\,\chi_{f^{\prime},m}^{\dagger}\}=0,\label{eq:CC_relation_2}
\end{equation}
\begin{equation}
[U_{n},\,L_{m}]=\delta_{nm}U_{n}.
\end{equation}
Note that the roles of the staggered fermion operators depend on the site $n$:
\begin{equation}
\chi_{f,n}^{\dagger}:\begin{cases}
\textrm{creation op. of particle} & n:\textrm{even}\\
\textrm{annihilation op. of anti-particle} & n:\textrm{odd}
\end{cases},\label{eq:particle_creation}
\end{equation}
\begin{equation}
\chi_{f,n}:\begin{cases}
\textrm{annihilation op. of particle} & n:\textrm{even}\\
\textrm{creation op. of anti-particle} & n:\textrm{odd}
\end{cases}.\label{eq:particle_annihilation}
\end{equation}
Thus, the operator $\chi_{f,n}^{\dagger}\chi_{f,n}$ counts the number of particles
on the even sites, whereas $\chi_{f,n}\chi_{f,n}^{\dagger}$
counts the number of anti-particles on the odd sites.
Considering that the particle has an electric charge of $+1$
and the anti-particle has $-1$, the charge density operator
at the site $n$ is given by
\begin{equation}
\rho_{f,n}=\chi_{f,n}^{\dagger}\chi_{f,n}+\frac{(-1)^{n}-1}{2}=\begin{cases}
\chi_{f,n}^{\dagger}\chi_{f,n} & n:\textrm{even}\\
-\chi_{f,n}\chi_{f,n}^{\dagger} & n:\textrm{odd}
\end{cases}.\label{eq:charge_density}
\end{equation}

In this work, we choose the open boundary condition in order to eliminate
the bosonic degrees of freedom having an infinite dimensional Hilbert space. The Gauss law (\ref{eq:Gauss_law_continuum}) is also discretized as
\begin{equation}
L_{n}-L_{n-1}=\sum_{f=1}^{N_{f}}\rho_{f,n},\label{eq:Gauss_law_lattice}
\end{equation}
where the left-hand side corresponds to the divergence of the electric
field and the right-hand side is the charge density (\ref{eq:charge_density}).

We set the explicit form of the (1+1)d gamma matrices
$\gamma^{0}=\sigma^{3}$, $\gamma^{1}=i\sigma^{2}$
and $\gamma^{5}=\gamma^{0}\gamma^{1}=\sigma^{1}$.
Using the operators $\chi_{f,n}$, $U_{n}$, and $L_{n}$ introduced above,
the lattice Hamiltonian is given by 
\cite{Funcke:2023lli,Dempsey:2023gib}
\begin{align}
H & =J\sum_{n=0}^{N-2}\left(L_{n}+\frac{\theta}{2\pi}\right)^{2}\nonumber \\
 & +\sum_{f=1}^{N_{f}}\left[-iw\sum_{n=0}^{N-2}\left(\chi_{f,n}^{\dagger}U_{n}\chi_{f,n+1}-\chi_{f,n+1}^{\dagger}U_{n}^{\dagger}\chi_{f,n}\right)+m_{\mathrm{lat}}\sum_{n=0}^{N-1}(-1)^{n}\chi_{f,n}^{\dagger}\chi_{f,n}\right],
\end{align}
where $J=g^{2}a/2$ and $w=1/2a$. Here we replace the mass $m$ of
the continuum theory by 
\begin{equation}
m_{\mathrm{lat}}:=m-\frac{N_{f}g^{2}a}{8}
\label{eq:latticemass}
\end{equation}
in the lattice Hamiltonian, following the recent proposal
\cite{Dempsey:2022nys} for eliminating $O(a)$ correction.
In the continuum theory, the chiral limit $m=0$ has the continuous chiral symmetry, and it contains $[\SU(N_f)_V/\mathbb{Z}_{N_f}]\times (\mathbb{Z}_{N_f})_L$ as a subgroup. 
With the above replacement, the lattice theory at $m=0$ maintains the discrete chiral symmetry $\mathbb{Z}_2\subset (\mathbb{Z}_{N_f})_L$ for even $N_f$, and this is the point protected by the remnant of the chiral symmetry. 

By adding up the lattice Gauss law equation (\ref{eq:Gauss_law_lattice})
from the boundary $n=0$ to the site $n$, we find that $L_{n}$ can be replaced by
\begin{align}
L_{n} & =L_{-1}+\sum_{f=1}^{N_{f}}\sum_{k=0}^{n}\rho_{f,k}\nonumber \\
 & =\sum_{f=1}^{N_{f}}\sum_{k=0}^{n}\chi_{f,k}^{\dagger}\chi_{f,k}
 +\frac{N_{f}}{2}\left(\frac{(-1)^{n}-1}{2}-n\right),
\end{align}
where we set $L_{-1}=0$.
Furthermore, we can set $U_{n}=1$ since the degrees of freedom of
$U_{n}$ can be absorbed by the U(1) phase of $\chi_{n}$. Then the
lattice Hamiltonian is written only by the fermions as $H=H_J+H_w+H_m$,
where the gauge part $H_J$ is given by
\begin{equation}
H_J = J\sum_{n=0}^{N-2}\left[
\sum_{f=1}^{N_{f}}\sum_{k=0}^{n}\chi_{f,k}^{\dagger}\chi_{f,k}
+\frac{N_{f}}{2}\left(\frac{(-1)^{n}-1}{2}-n\right)+\frac{\theta}{2\pi}\right]^{2},
\label{eq:H_J}
\end{equation}
and the kinetic term $H_w$ and the mass term $H_m$ of the fermions are
\begin{equation}
H_w = -iw\sum_{f=1}^{N_{f}}\sum_{n=0}^{N-2}
\left(\chi_{f,n}^{\dagger}\chi_{f,n+1}-\chi_{f,n+1}^{\dagger}\chi_{f,n}\right),
\label{eq:H_w}
\end{equation}
\begin{equation}
H_m = m_{\mathrm{lat}}\sum_{f=1}^{N_{f}}\sum_{n=0}^{N-1}(-1)^{n}\chi_{f,n}^{\dagger}\chi_{f,n}.
\label{eq:H_m}
\end{equation}

\subsection{Map to the spin system}
\label{subsec:Map_to_spin}

Now, we map the Hamiltonian written by the staggered fermions to the spin Hamiltonian. Such a spin Hamiltonian formalism is useful to apply tensor network methods and quantum computations.

The $N_{f}\times N$ degrees of freedom of the staggered fermion $\chi_{f,n}$
can be described by the same number of spin-1/2 degrees of freedom.
The Hilbert space of such a spin system is given by 
\begin{equation}
\mathcal{H}=\bigotimes_{f=1}^{N_{f}}\bigotimes_{n=0}^{N-1}\mathcal{H}_{f,n},\label{eq:H_space}
\end{equation}
where $\mathcal{H}_{f,n}$ is the local Hilbert space of the single spin-1/2 state.
A general state $\ket{\Psi}$ in this Hilbert space can be described by
a superposition of all possible spin configurations $\boldsymbol{s}$,
\begin{equation}
\ket{\Psi}=\sum_{\boldsymbol{s}}\Psi(\boldsymbol{s})\ket{\boldsymbol{s}},\label{eq:general_state}
\end{equation}
\begin{equation}
\ket{\boldsymbol{s}}\in\left\{ \left.\bigotimes_{f=1}^{N_{f}}\bigotimes_{n=0}^{N-1}\ket{s_{f,n}}_{f,n}\right|\ket{s_{f,n}}_{f,n}=\ket{\uparrow},\ket{\downarrow}\right\} .
\end{equation}
The spin up $\ket{\uparrow}$ and down $\ket{\downarrow}$ state are the eigenstates
of the Pauli matrix $\sigma^{z}$ with the eigenvalues $+1$ and $-1$, respectively. 

The map to the spin system can be achieved by the so-called Jordan-Wigner transformation.
The fermion operators $\chi_{f,n}$ for the two flavors $f=1,2$ are represented
by spin operators as follows:
\begin{align}
\chi_{1,n} & =\sigma_{1,n}^{-}\prod_{j=0}^{n-1}(-\sigma_{2,j}^{z}\sigma_{1,j}^{z}), & \chi_{1,n}^{\dagger} & =\sigma_{1,n}^{+}\prod_{j=0}^{n-1}(-\sigma_{2,j}^{z}\sigma_{1,j}^{z}),\label{eq:JW_1}\\
\chi_{2,n} & =\sigma_{2,n}^{-}(-i\sigma_{1,n}^{z})\prod_{j=0}^{n-1}(-\sigma_{2,j}^{z}\sigma_{1,j}^{z}), & \chi_{2,n}^{\dagger} & =\sigma_{2,n}^{+}(i\sigma_{1,n}^{z})\prod_{j=0}^{n-1}(-\sigma_{2,j}^{z}\sigma_{1,j}^{z}),\label{eq:JW_2}
\end{align}
where we define
\begin{equation}
\sigma_{f,n}^{\pm}=\frac{1}{2}(\sigma_{f,n}^{x}\pm i\sigma_{f,n}^{y}).
\end{equation}
The Pauli matrices $\sigma_{f,n}^{a}$ ($a=x,y,z$) act on the spin $\ket{s_{f,n}}_{f,n}$
at the site $n$ of the flavor $f$.
They do not commute only if they are on the same site of the same flavor, so that
\begin{equation}
\left[\sigma_{f,n}^{a},\,\sigma_{f^{\prime},n^{\prime}}^{b}\right]=2i\delta_{ff^{\prime}}\delta_{nn^{\prime}}\epsilon^{abc}\sigma_{f,n}^{c}.
\end{equation}
We can check that the canonical anti-commutation relations (\ref{eq:CC_relation_1})
and (\ref{eq:CC_relation_2}) are satisfied thanks to the properties of the Pauli matrices. 

Note that this is not a unique way of translation to the spin system
which realizes the anti-commutation relations.
Different transformations give different representations of the original Hamiltonian.
We choose this transformation since various local operators can be constructed by
only a few numbers of the Pauli matrices.
The spin representation of the Hamiltonian and the observables defined above
are summarized in Appendix \ref{sec:Spin_op}.

\subsection{Local observables}
\label{subsec:local_obs}

Let us consider the meson operators (\ref{eq:pi_meson}) -- (\ref{eq:eta_meson}) on the lattice. 
Based on the continuum descriptions,
it is natural to define the lattice version of these operators as follows:
\begin{align}
\pi(n) & :=PS_{1,n}-PS_{2,n}, \label{eq:pi_op} \\
\eta(n) & :=PS_{1,n}+PS_{2,n}, \label{eq:eta_op} \\
\sigma(n) & :=S_{1,n}+S_{2,n}. \label{eq:sigma_op}
\end{align}
Here $S_{f,n}$ and $PS_{f,n}$ are the scalar and pseudo-scalar operators
for the flavor $f=1,2$ on the lattice, respectively.
In order to obtain their explicit form, we rewrite the scalar condensate
$(\bar{\psi}\psi)_{f}$ by the staggered fermion, so that
\begin{align}
(\bar{\psi}\psi)_{f} & =(\psi_{u}^{\dagger}\psi_{u}-\psi_{d}^{\dagger}\psi_{d})_{f}, \nonumber \\
 & =\frac{1}{2a}(-1)^{n}(\chi_{f,n}^{\dagger}\chi_{f,n}-\chi_{f,n+1}^{\dagger}\chi_{f,n+1}).\label{eq:scalar_cond_original}
\end{align}
Similarly, the pseudo-scalar condensate $(\bar{\psi}\gamma^{5}\psi)_{f}$ is given by 
\begin{align}
(\bar{\psi}\gamma^{5}\psi)_{f} & =(\psi_{u}^{\dagger}\psi_{d}-\psi_{d}^{\dagger}\psi_{u})_{f},\nonumber \\
 & =\frac{1}{2a}(-1)^{n}(\chi_{f,n}^{\dagger}\chi_{f,n+1}-\chi_{f,n+1}^{\dagger}\chi_{f,n}).\label{eq:p-scalar_cond_original}
\end{align}
These operators have a site-by-site fluctuation due to the staggered fermion.
Here we define the lattice scalar condensate operator
by the two-site average of (\ref{eq:scalar_cond_original}), namely
\begin{align}
S_{f}(n) & :=\frac{1}{2}\left[(\bar{\psi}\psi)_{f,n-1}+(\bar{\psi}\psi)_{f,n}\right], \nonumber \\
 & =\frac{1}{4a}(-1)^{n}(-\chi_{f,n-1}^{\dagger}\chi_{f,n-1}+2\chi_{f,n}^{\dagger}\chi_{f,n}-\chi_{f,n+1}^{\dagger}\chi_{f,n+1}),\label{eq:S_staggered}
\end{align}
for $n=1,2,\cdots,N-2$.
The lattice pseudo-scalar condensate operator is also defined by
the two-site average of (\ref{eq:p-scalar_cond_original}) with a factor $-i$,
\begin{align}
PS_{f}(n) & :=-\frac{i}{2}\left[(\bar{\psi}\gamma^{5}\psi)_{f,n-1}+(\bar{\psi}\gamma^{5}\psi)_{f,n}\right], \nonumber \\
 & =\frac{i}{4a}(-1)^{n}(\chi_{f,n-1}^{\dagger}\chi_{f,n}-\chi_{f,n}^{\dagger}\chi_{f,n-1}-\chi_{f,n}^{\dagger}\chi_{f,n+1}+\chi_{f,n+1}^{\dagger}\chi_{f,n}),\label{eq:PS_staggered}
\end{align}
for $n=1,2,\cdots,N-2$. Note that both of $S_{f}(n)$ and $PS_{f}(n)$
are composed of the staggered fermions at the three sites $n$ and $n\pm1$.

\subsection{Global observables}
\label{subsec:global_obs}

We will define the quantum number ($J_z, \bm{J}^2, C$, and $P$) and momentum operators, which will be useful to distinguish the eigenstates of the Hamiltonian. 
These operators can be described by some global observables, which act on the whole lattice.

First of all, let us focus on the isospin operator (\ref{eq:J_cont}).
We define the lattice version in terms of the staggered fermion.
The isospin $J_{z}$ operator counts the number of particles of each flavor
with the factor $\pm1/2$ on even sites and the number of anti-particles
with the opposite sign on odd sites. Thus, it can be realized by 
\begin{equation}
J_{z}=\frac{1}{2}\sum_{n=0}^{N-1}
\left(\chi_{1,n}^{\dagger}\chi_{1,n}-\chi_{2,n}^{\dagger}\chi_{2,n}\right).
\label{eq:J_z}
\end{equation}
It is convenient to define the isospin $J_{\pm}$ operators by
\begin{equation}
J_{\pm}=J_{x} \pm i J_{y}.
\end{equation}
Based on the role of fermion operators (\ref{eq:particle_creation})
and (\ref{eq:particle_annihilation}), $J_{+}$ operator is given by 
\begin{equation}
J_{+}=\sum_{n=0}^{N-1}\chi_{1,n}^{\dagger}\chi_{2,n},
\label{eq:J_plus}
\end{equation}
which transforms $f=2$ particle to $f=1$ particle on even sites
and $f=1$ anti-particle to $f=2$ anti-particle on odd sites.
Similarly, $J_{-}$ operator is given by 
\begin{equation}
J_{-}=\sum_{n=0}^{N-1}\chi_{2,n}^{\dagger}\chi_{1,n},
\label{eq:J_minus}
\end{equation}
which transforms $f=1$ particle to $f=2$ particle on even sites
and $f=2$ anti-particle to $f=1$ anti-particle on odd sites.
Then the Casimir operator $\bm{J}^{2}$ can also be defined
as the combination of the operators above by
\begin{equation}
\bm{J}^{2} = \frac{1}{2}(J_{+}J_{-}+J_{+}J_{-})+J_{z}^{2}.
\label{eq:J2}
\end{equation}

Second, we will consider the charge conjugation and parity operators.
For this purpose, let us discuss the description of the particle and anti-particle
as a spin state. Applying the Jordan-Winger transformation, the spin representation of
the charge density operator (\ref{eq:charge_density}) is given by
\begin{equation}
\rho_{f,n}=\frac{\sigma_{f,n}^{z}+1}{2}+\frac{(-1)^{n}-1}{2}=
\begin{cases}
(\sigma_{f,n}^{z}+1)/2 & n:\textrm{even},\\
(\sigma_{f,n}^{z}-1)/2 & n:\textrm{odd}.
\end{cases}
\end{equation}
This operator counts the number of particles with $+1$ on even sites
and the number of anti-particles with $-1$ on odd sites.
We can confirm that the particle is described by the spin-up state $\ket{\uparrow}$
on even sites by taking the expectation value
\begin{equation}
\bra{\uparrow}\rho_{f,n}\ket{\uparrow}_{f,n}=\begin{cases}
1 & n:\textrm{even},\\
0 & n:\textrm{odd}.
\end{cases}\label{eq:rho_up}
\end{equation}
Similarly, we find that the anti-particle is described by
the spin-down state $\ket{\downarrow}$ on odd sites as
\begin{equation}
\bra{\downarrow}\rho_{f,n}\ket{\downarrow}_{f,n}=\begin{cases}
0 & n:\textrm{even},\\
-1 & n:\textrm{odd}.
\end{cases}\label{eq:rho_down}
\end{equation}

Based on this fact, charge conjugation, namely the exchange of particles
and anti-particles can be performed by the exchange of even sites and odd sites.
In addition, the spin-up state should be replaced by the spin-down state, and vice versa.
These operations can be realized by the 1-site translation of the lattice
and the multiplication of $\sigma^{x}$ operators.
Thus, we define the charge conjugation operator by~\cite{Banuls:2013jaa} 
\begin{equation}
C:=\prod_{f=1}^{N_{f}}\left(\prod_{n=0}^{N-1}\sigma_{f,n}^{x}\right)\left(\prod_{n=0}^{N-2}(\mathrm{SWAP})_{f;N-2-n,N-1-n}\right),\label{eq:C_spin}
\end{equation}
where the swap operator is given by
\begin{equation}
(\mathrm{SWAP})_{f;j,k}=\frac{1}{2}\left(\bm{1}_{f,j}\bm{1}_{f,k}+\sum_{a}\sigma_{f,j}^{a}\sigma_{f,k}^{a}\right),
\end{equation}
using the Pauli matrices.
As the name suggests, the swap operator exchanges
the state $\ket{s}_{f,j}$ and $\ket{s^{\prime}}_{f,k}$, namely
\begin{equation}
(\mathrm{SWAP})_{f;j,k}\ket{s}_{f,j}\otimes\ket{s^{\prime}}_{f,k}=\ket{s^{\prime}}_{f,j}\otimes\ket{s}_{f,k}.
\end{equation}
The product of the swap operators in (\ref{eq:C_spin}) realizes the 1-site translation.
The charge conjugation defined in this way satisfies $C^{\dagger}C=1$,
but $C^{2}\neq1$. 
When we take the periodic boundary condition, $C^2=1$ is achieved in the continuum limit, but this is not the case for the open boundary condition.
Moreover, the Hamiltonian does not commute with $C$ due to the presence of the boundaries, 
and we will actually see the eigenstates of the Hamiltonian give $|\Braket{C}|<1$. 
Therefore, $C$ does not give a good quantum number when we take the staggered-fermion regularization with the open boundary condition. 
In this study, following the observation of Ref.~\cite{Banuls:2013jaa}, we assume that the sign of $\mathrm{Re}\Braket{C}$ remembers the original sign of $C$ for each eigenstate.
We discuss this prescription in detail in Appendix~\ref{sec:C_test}.

Next, we define the parity operator. The parity transformation $x\rightarrow-x$
can be achieved by flipping the order of the lattice sites.
The site $n\in\{0,1,\cdots,N-1\}$ is mapped to the site $n^{\prime}=N-1-n$.
However, this operation also exchanges particles and anti-particles
since the roles of even sites and odd sites are exchanged when $N$ is even.
Thus, an additional operation of 1-site translation is necessary to fix it.
We define the parity operator by
\begin{align}
P:=\prod_{f=1}^{N_{f}} & \left(\prod_{j=0}^{N/2-1}\sigma_{f,2j+1}^{z}\right)\nonumber \\
\times & \left(\prod_{n=0}^{N-2}(\mathrm{SWAP})_{f;N-2-n,N-1-n}\right)\left(\prod_{n=0}^{N/2-1}(\mathrm{SWAP})_{f;n,N-1-n}\right),\label{eq:P_spin}
\end{align}
where the products of the swap operators perform the reversal $n\rightarrow n^{\prime}$
and the 1-site translation.\footnote{
If we implement the reversal $n\rightarrow n^{\prime}$ in this manner,
the bond dimension of MPO grows exponentially with $N$.
Thus, in practice, we apply the reversal by transposing all the matrices in MPS.}
The additional factors of $\sigma^{z}$ come from the shift of the staggered phase,
which corresponds to $\gamma^{0}$ in the parity transformation
of the Dirac fermion $\psi(x)\rightarrow\gamma^{0}\psi(-x)$.
As we mentioned for the $C$ operator, the $P$ operator in the open boundary condition does not commute with the Hamiltonian as it contains the $1$-unit lattice translation. 
Therefore, we take the same prescription to determine the parity quantum number for each state as in the case of $C$.

Finally, the other important quantity is a total momentum operator, which can be used to
identify the momentum excitation~\cite{Banuls:2013jaa}.
We start with the continuum description of the gauge invariant operator,
\begin{equation}
K=\sum_{f=1}^{N_{f}}\int dx\,\psi_{f}^{\dagger}(i\partial_{x}-A_1)\psi_{f},
\end{equation}
which commutes with the continuum Hamiltonian \eqref{eq:H_cont} under the periodic boundary condition using the Gauss-law constraint \eqref{eq:Gauss_law_continuum}.
In our case with the open boundary condition, it does not commute with the Hamiltonian since the translational symmetry is explicitly broken.
Thus, the expectation value $\Braket{K}$ is no longer the quantum number in the strictest sense.
However, we will see that the operator is still useful as an approximate one
to investigate the mass spectrum of the model.

Let us consider its lattice version.
Here we set $A_1(x)=0$ since we fix the gauge $U_n=1$ in our setup.
The combination $\psi_{f}^{\dagger}\partial_{x}\psi_{f}$ of the Dirac fermion corresponds to 
\begin{equation}
\psi_{f}^{\dagger}\partial_{x}\psi_{f}
=(\psi_{u}^{\dagger}\partial_{x}\psi_{u}+\psi_{d}^{\dagger}\partial_{x}\psi_{d})_{f}
=\frac{1}{2a}\chi_{f,n}^{\dagger}(\chi_{f,n+2}-\chi_{f,n}),
\end{equation}
in terms of the staggered fermion. There is another possible combination
\begin{equation}
-(\partial_{x}\psi_{f}^{\dagger})\psi_{f}
=-\frac{1}{2a}(\chi_{f,n+2}^{\dagger}-\chi_{f,n}^{\dagger})\chi_{f,n},
\end{equation}
given by the integral by parts ignoring boundary term.
Then we define the total momentum on the lattice as a Hermitian operator by
taking symmetric combination
\begin{align}
K & :=\frac{i}{2}\sum_{f=1}^{N_{f}}\sum_{n=0}^{N-3}\frac{1}{2a}\left[\chi_{f,n}^{\dagger}(\chi_{f,n+2}-\chi_{f,n})-(\chi_{f,n+2}^{\dagger}-\chi_{f,n}^{\dagger})\chi_{f,n}\right], \nonumber \\
 & =\frac{i}{4a}\sum_{f=1}^{N_{f}}\sum_{n=1}^{N-2}(\chi_{f,n-1}^{\dagger}\chi_{f,n+1}-\chi_{f,n+1}^{\dagger}\chi_{f,n-1}).\label{eq:K_staggered}
\end{align}
This operator does not commute with the term $H_w$ (\ref{eq:H_w}) and $H_J$ (\ref{eq:H_J}) of the lattice Hamiltonian due to the open boundary.
We also note that the latter $[K,H_J]$ has an $O(a)$ violation effect even in the periodic boundary condition.

\section{Calculation method and the simulation setup}
\label{sec:dmrg}

We employ the density-matrix renormalization group (DMRG)
\cite{White:1992zz,White:1993zza,Schollw_ck_2005,Schollw_ck_2011} to study
the spin Hamiltonian of the 2-flavor Schwinger model after the Jordan-Winger transformation,
whose explicit form is given by \eqref{eq:H_spin}.
The DMRG is known as an efficient method to study (1+1)d gapped spin systems
and has been developed mainly in the field of condensed matter physics. 
We utilized the C++ library of ITensor \cite{itensor}
to perform the tensor network calculation of this work.
Let us briefly explain the basic idea of DMRG to obtain the ground state and excited states, and then we explain the details of parameter settings.

\subsection{Quick review of DMRG}
\label{sec:method}

In the spin systems, any wave function $\ket{\Psi}$ can be expressed as the form of the matrix product states (MPS),
\begin{equation}
    \ket{\Psi} =\sum_{i_1,\ldots, i_N=1}^{2}\tr[A_1^{(i_1)}\cdots A_{N}^{(i_N)}]\ket{i_1\ldots i_N}, 
\end{equation}
by repeating the singular-value decomposition (SVD). Here, $i_n=1,2$ denotes the spin degrees of freedom at the $n$-th site, $A_n^{(i_n)}$ denotes a $D\times D$ matrix, and this size $D$ is called the bond dimension. 
The upper bound for the entanglement entropy of $\ket{\Psi}$ is given by $\ln D$. 
Therefore, the MPS gives a useful tool to study the many-body states with low entanglement entropies, such as the ground state of the $(1+1)$d gapped systems~\cite{Stoudenmire_2012,Wall_2012}. 
For the 2-flavor Schwinger model, we arrange the site index $n$ and the flavor index $f$ on the 1d lattice with the single index to apply DMRG.
The ordering of the indices is chosen so that the behavior of entanglement entropy is reproduced appropriately with a reasonable bond dimension.
This point is discussed in Appendix~\ref{sec:order_of_index}.

The DMRG is a variational algorithm based on the MPS. 
In each step of the algorithm, the matrices are updated
to decrease the energy $E=\Braket{\Psi|H|\Psi}$ as a cost function. In addition, we perform the low-rank approximation and thus
the smaller singular values are discarded, which amount to an error $\Delta$.
We determine the bound dimension by setting the maximal bond dimension and also the cutoff parameter $\varepsilon$
on the error so that $\Delta\le\varepsilon$.
Smaller $\varepsilon$ gives a better approximation, but it also requires a larger bond dimension and increases the computational costs. 
We can also effectively calculate the expectation values or correlation functions of local operators by rewriting those operators in the form of matrix product operators (MPOs)
and then taking contractions with the ground state $\ket{\Psi}$.\footnote{
The bond dimension of the MPO is determined similarly by a cutoff parameter
$\varepsilon$ in the SVD.
We set $\varepsilon=10^{-13}$ for MPOs, which is sufficiently small
so that the bond dimension of the MPO is saturated.}

We can use DMRG to obtain the low-energy excited states in a recursive way. 
Assume that we already find the energy eigenstates $\ket{\Psi_{\ell'}}$ with $\ell'=0,1,\ldots, \ell-1$ from below. 
Then, we apply the same technique to find the $\ell$-th state $\ket{\Psi_\ell}$ by changing the Hamiltonian for the cost function as  
\begin{equation}
H_{\ell}=H+W\sum_{\ell^{\prime}=0}^{\ell-1}\ket{\Psi_{\ell^{\prime}}}\bra{\Psi_{\ell^{\prime}}},  
\end{equation}
where $W>0$ is a weight to impose the orthogonality.
We can generate the excited states from the ground state to any level step by step.

\subsection{Simulation setup}
\label{sec:setup}

Let us explain our parameter setup when using the ITensor~\cite{itensor}. 
The gauge coupling $g$ has mass dimension $1$ in $1+1$d QED, and thus we can measure the energy scale in the unit of $g$ by setting $g=1$. 
In this work, we always set $g=1$ and the fermion mass $m=0.1$, so the photon mass is $\mu=\sqrt{\frac{2}{\pi}}\simeq 0.8$. The lattice fermion mass~\eqref{eq:latticemass} becomes $m_{\mathrm{lat}}=0.1-\frac{a}{4}$.
The theta angle is normally set to $\theta = 0$,
except when measuring the one-point function of the pion at $\theta = 2\pi$.

For the correlation-function scheme and the one-point-function scheme,
we use the lattice size of $N=160$.
The lattice spacing is set to $a\approx 0.25$ so that the physical size is $L=a(N-1)=39.8$.
The number of DMRG steps called the sweeps, is set to $N_{\mathrm{sweep}}=20$.
We generate the ground state for four different values of the cutoff parameter: $\varepsilon = 10^{-10}$, $10^{-12}$, $10^{-14}$, and $10^{-16}$.
To characterize the bond dimension of the MPS, we focus on the largest number of nonzero singular values,
which we call the effective bond dimension, denoted as $D_{\mathrm{eff}}$. 
In our computations, we set the maximal bond dimension large enough so that $D_{\mathrm{eff}}$ is solely controlled by the cutoff $\varepsilon$ for the above physical setup. 
We observe $D_{\mathrm{eff}}$ to be about 400, 800, 1600, and 2800
for the respective values of $\varepsilon$ above.

For the dispersion-relation scheme,
we generate many excited states up to $\ell=23$, which require a lot of computational cost.
Therefore, we choose a smaller lattice size of $L=19.8$ with $N=100$ and $a=0.2$.
The excited states are generated
with a cutoff of $\varepsilon = 10^{-10}$ and a weight parameter of $W=10$.
To achieve better convergence of higher states,
we increase the number of sweeps to $N_{\mathrm{sweep}}=50$.
The bond dimension is about 500 for the ground state while it is at most 2300 for the excited states.

As an initial state of the DMRG, we choose the N\'{e}el state, which is
a direct product of the spin-down states on even sites and the spin-up state on odd sites,
\begin{equation}
\Ket{\mathrm{N\acute{e}el}}=\bigotimes_{f=1}^{N_{f}}
\ket{\downarrow}_{f,0}\ket{\uparrow}_{f,1}\cdots\ket{\downarrow}_{f,N-2}\ket{\uparrow}_{f,N-1}.
\end{equation}
Based on (\ref{eq:rho_up}) and (\ref{eq:rho_down}), the N\'{e}el states
is regarded as a zero-particle state.
We also impose the charge conservation condition during the DMRG,
so that the MPS satisfies the condition $Q=0$, where
\begin{equation}
Q=\sum_{f=1}^{N_{f}}\sum_{n=0}^{N-1}\rho_{f,n}=\frac{1}{2}\sum_{f=1}^{N_{f}}\sum_{n=0}^{N-1}\sigma_{f,n}^{z}.
\end{equation}
We note that the Gauss law with the usual open boundary on both sides requires $Q=0$ on the physical states.

\section{Simulation results}
\label{sec:result}

In this section, we explain our numerical results for the meson spectrum of the 2-flavor Schwinger model at $\theta=0$. 
We apply three distinct methods in our computations of the meson spectrum:
\begin{itemize}
    \item the correlation-function scheme
    \item the one-point-function scheme 
    \item the dispersion-relation scheme
\end{itemize}
Each method has its own pros and cons, and we are going to discuss them. 
We will see that all these schemes give consistent results.

\subsection{Correlation-function scheme}
\label{subsec:result_CF}

In the relativistic quantum field theories, the Hilbert space only plays a secondary role,
and we are supposed to reconstruct all the physical information from the correlation functions. 
In the conventional Euclidean lattice gauge theory, people usually follow this dogma,
and the mass spectrum is obtained from the correlation function in the imaginary time direction.
We can take the same approach also in the Hamiltonian formalism by the measurement of the spatial correlation function with the distance $r=|x-y|$. 

\begin{figure}[tb]
\centering
\includegraphics[scale=0.45]{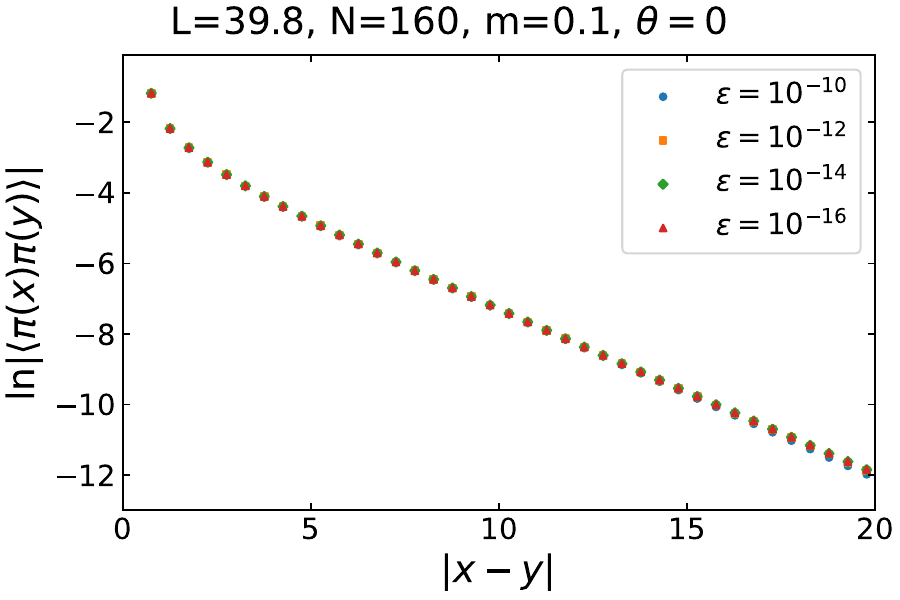}
\quad 
\includegraphics[scale=0.45]{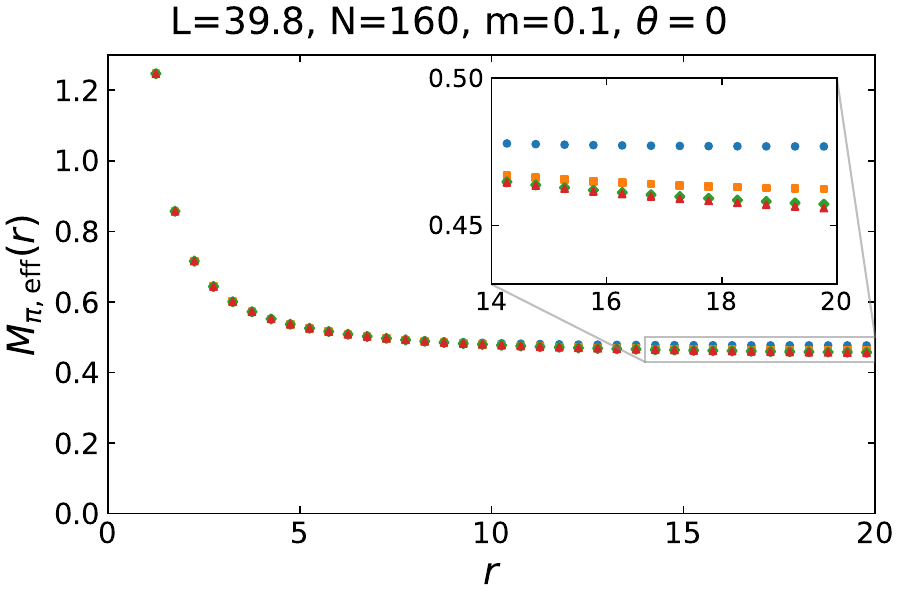}
\caption{\label{fig:cf_pi}
(Left) The correlation function of pion $\ln |\Braket{\pi(x)\pi(y)}|$ is plotted against the distance $r=|x-y|$ for various values of $\varepsilon$.
The number of lattice sites is $N=160$ and the lattice spacing $a$ is determined so that $L=a(N-1)=39.8$.
(Right) The effective mass of the pion $M_{\pi,\mathrm{eff}}(r)$ (3-point average) calculated
from the correlation function in the left panel is plotted against $r$.}
\end{figure}

First, let us work on pions, and we consider the equal-time spatial correlation function,
\begin{equation}
C_{\pi}(r)=\Braket{\pi(x)\pi(y)},
\end{equation}
where $\pi(x)$ denotes the operator defined by (\ref{eq:pi_op}) with $x=na$. 
In order to evade the boundary effect as much as possible, we compute $C_{\pi}(r)$ by changing $x$ and $y$ symmetrically as $x=(L-r)/2$ and $y=(L+r)/2$, 
and the range of $r$ is restricted to $0\le r\le L/2$.
The results are shown in the left panel of Fig.~\ref{fig:cf_pi} in the logarithmic scale, and the pion mass can be extracted from the exponential decay of $C_{\pi}(r)$.
Here, the data with different colors represent the different values of the cutoff parameter $\varepsilon$.

It is convenient to use the so-called effective mass defined by 
\begin{equation}
\tilde{M}_{\pi,\mathrm{eff}}(r)=-\frac{1}{2a}\log\frac{C_{\pi}(r+2a)}{C_{\pi}(r)},
\end{equation}
where $2a$ comes from the step size of changing $r$.
We further take the 3-point average of the effective mass
\begin{equation}
M_{\pi,\mathrm{eff}}(r)=\frac{1}{4}\tilde{M}_{\pi,\mathrm{eff}}(r-2a)+\frac{1}{2}\tilde{M}_{\pi,\mathrm{eff}}(r)+\frac{1}{4}\tilde{M}_{\pi,\mathrm{eff}}(r+2a)
\end{equation}
to suppress a remaining oscillation caused by the staggered fermion.
The result of $M_{\pi,\mathrm{eff}}(r)$ is shown in the right panel of Fig.~\ref{fig:cf_pi}. 

One might be tempted to think that the pion mass corresponds to the plateau value of $M_{\pi,\mathrm{eff}}(r)$, 
and the result with the DMRG cutoff $\varepsilon=10^{-10}$ actually seems to become constant for $r\gtrsim 10$ almost exactly. 
However, this is the fake plateau due to the low-rank approximation. 
The point is that the leading asymptotic behavior of the spatial correlator is not purely the exponential decay,
and it would take the Yukawa-type form asymptotically as $r\to \infty$, 
\begin{equation}
    C_{\pi}(r)\sim \frac{1}{r^{\alpha}}\exp(-M_{\pi} r). 
\end{equation}
We actually have $\alpha=1/2$ for the $(1+1)$d free massive boson, and we shall discuss the detailed analysis in Appendix~\ref{sec:CF_nf1} in the case of the $1$-flavor Schwinger model. 
As a result, the effective mass for the Yukawa-type correlation function is given by 
\begin{equation}
M_{\pi,\mathrm{eff}}(r)=-\frac{d}{dr}\log C_\pi(r)\sim\frac{\alpha}{r}+M_\pi, 
\end{equation}
and there must be an additional $O(1/r)$ contribution on top of the actual mass $M_\pi$. 

\begin{figure}[tb]
\centering
\includegraphics[scale=0.45]{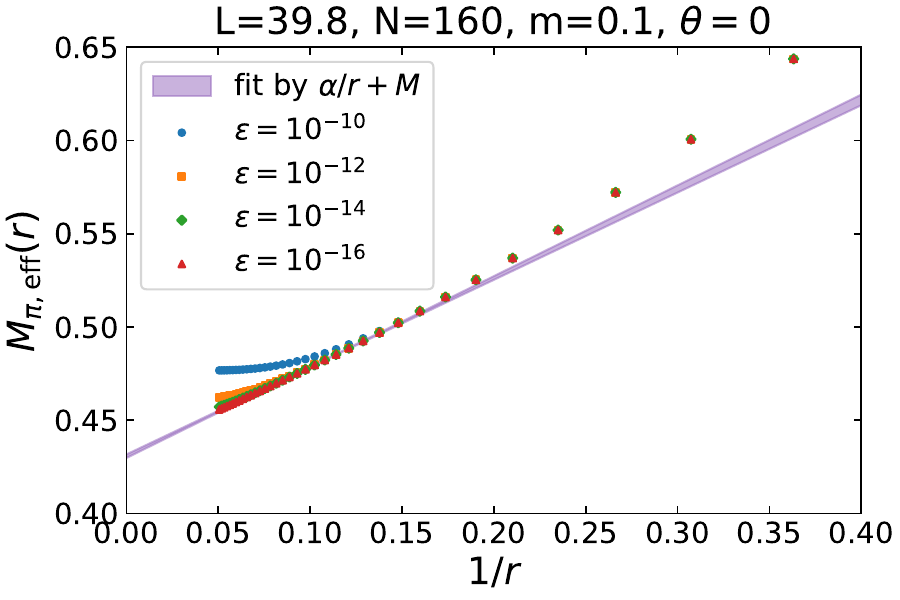}
\vspace{-1.2em}
\caption{\label{fig:fitMeff_pi}
The effective mass of the pion $M_{\pi,\mathrm{eff}}(r)$ is plotted against $1/r$.
The data points for $\varepsilon=10^{-16}$ are fitted by $\alpha/r+M$ inside the region
$0.075\protect\leq1/r\protect\leq0.15$.
The fitting result is depicted by the shaded band with systematic error.}
\end{figure}

Motivated by this fact, we plot $M_{\pi,\mathrm{eff}}(r)$ against $1/r$ in Fig.~\ref{fig:fitMeff_pi}.
We can see that the behavior of the effective mass strongly depends on the cutoff $\varepsilon$ especially when $r$ is large. 
When $\varepsilon$ is not small enough, we observe the saturation of $M_{\pi,\mathrm{eff}}(r)$, and then the $\alpha/r$ term seems to be absent. 
We note that the low-rank approximation of the DMRG is similar to the approximation of the transfer matrix by a finite matrix. 
Thus, $C_{\pi}(r)$ in the DMRG is approximated by the sum of purely exponential functions, and we need sufficiently large bond dimensions to reproduce the non-exponential corrections, such as $1/r^{\alpha}$. 

In fact, we can observe in Fig.~\ref{fig:fitMeff_pi} that the development of the $1/r$-behavior in $M_{\pi,\mathrm{eff}}(r)$ for large $r$
by making $\varepsilon$ sufficiently small, i.e. the bond dimension sufficiently large.
We estimate the mass $M_\pi$ by the linear extrapolation $1/r\rightarrow 0$
of the result for the largest bond dimension with $\varepsilon=10^{-16}$,
which is performed by fitting the data points with $\alpha/r+M_\pi$.
To evaluate the systematic errors from the uncertainty of the fitting range,
we try a lot of fittings by changing the fitting range inside the region $0.075\leq1/r\leq0.15$, 
and we obtain the probability distribution of the fitting results.
The best-fitting result and its error are estimated from the position and the width of the peak, respectively.
Thus, we obtained
\begin{equation}
    M_\pi=0.431(1),
\end{equation}
with $\alpha=0.477(9)$, and the fitting lines are drawn as the purple shadow in Fig.~\ref{fig:fitMeff_pi}.

\begin{figure}[tb]
\centering
\includegraphics[scale=0.45]{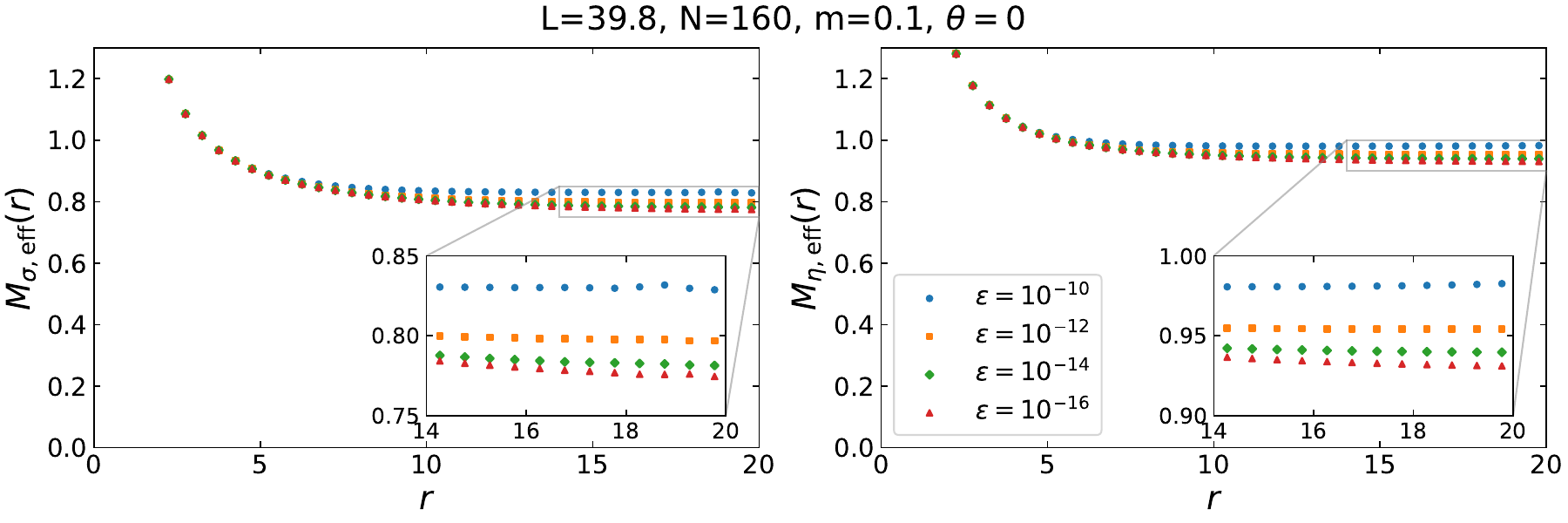}
\caption{\label{fig:Meff_sigma-eta}
The effective mass of sigma meson $M_{\sigma,\mathrm{eff}}(r)$ (left)
and eta meson $M_{\eta,\mathrm{eff}}(r)$. }
\end{figure}

\begin{figure}[tb]
\centering
\includegraphics[scale=0.45]{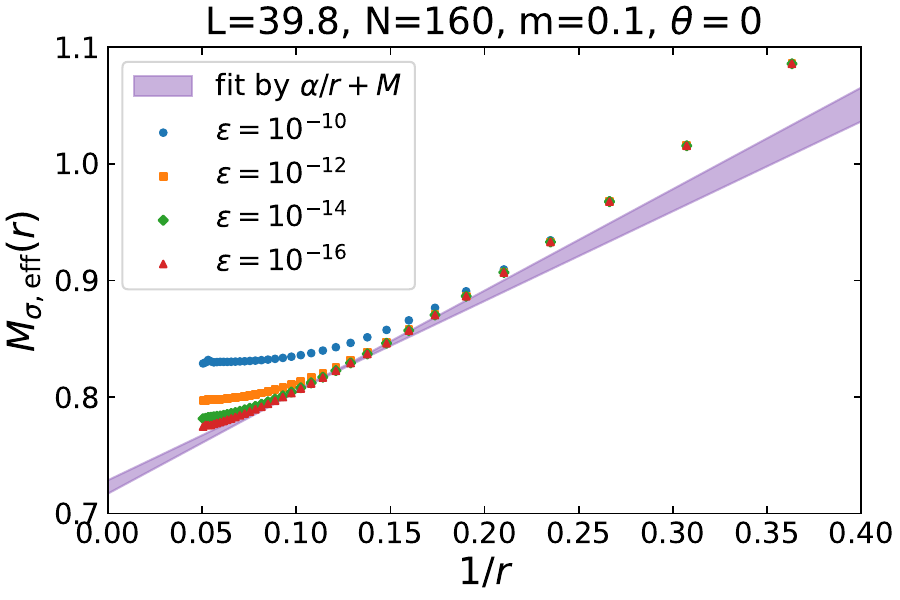}
\includegraphics[scale=0.45]{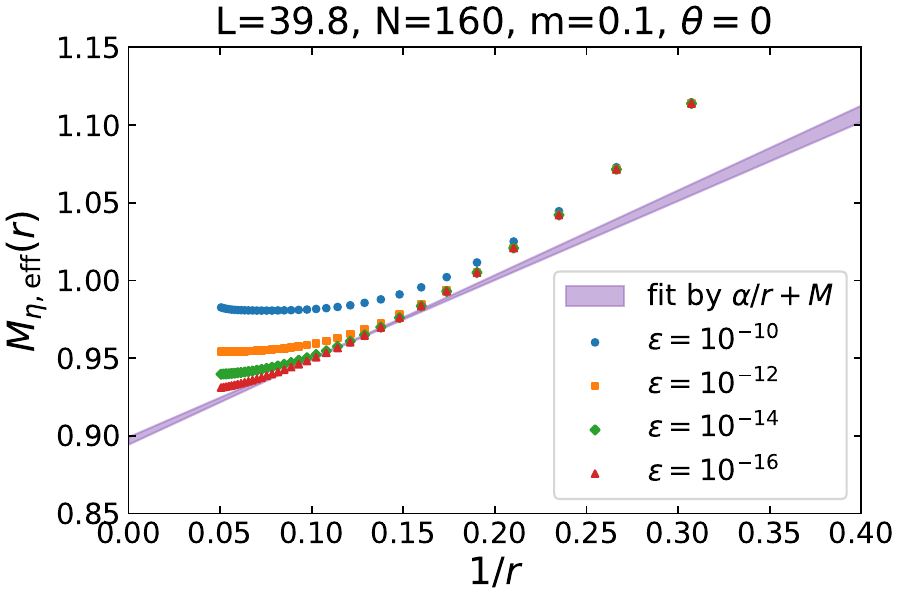}
\caption{\label{fig:fitMeff_sigma-eta}
The effective mass of sigma meson $M_{\sigma,\mathrm{eff}}(r)$ (left)
and eta meson $M_{\eta,\mathrm{eff}}(r)$ (right) are plotted against $1/r$.
The data points for $\varepsilon=10^{-16}$ are fitted by $\alpha/r+M$
inside the range $0.075\protect\leq1/r\protect\leq0.15$.
The fitting results are depicted by the shaded bands with systematic errors.}
\end{figure}

Next, we perform similar analyses for sigma meson (\ref{eq:sigma_op}) and eta meson (\ref{eq:eta_op}). 
Since these are isospin singlets, their one-point functions are not zero, and we subtract the disconnected parts from the correlation functions, 
\begin{align}
    C_{\sigma}(r)&=
    \langle \sigma(x)\sigma(y)\rangle-\langle \sigma (x)\rangle \langle \sigma(y)\rangle, \\
    C_{\eta}(r)&=
    \langle \eta(x)\eta(y)\rangle-\langle \eta(x)\rangle \langle \eta(y)\rangle, 
\end{align}
with $x=(L-r)/2$, $y=(L+r)/2$. 
We then compute the 3-point averages of the effective mass,
$M_{\sigma,\mathrm{eff}}(r)$ and $M_{\eta,\mathrm{eff}}(r)$,
and they are shown in Fig.~\ref{fig:Meff_sigma-eta}.
The difference in the asymptotic behavior is observed by changing $\varepsilon$
also in these cases.
We plot the effective masses of the sigma and eta mesons against $1/r$ in
Fig.~\ref{fig:fitMeff_sigma-eta} to see the asymptotic behavior. 
They approach $\propto 1/r$ for smaller $\varepsilon$ as expected.
We fit the data for $\varepsilon=10^{-16}$ by $\alpha/r+M$ inside the region $0.075\leq1/r\leq0.15$ and estimate the systematic error. 
Then we obtained 
\begin{equation}
    M_\sigma = 0.722(6),
\end{equation}
with $\alpha=0.83(5)$ for sigma meson, and 
\begin{equation}
    M_\eta=0.899(2),
\end{equation}
with $\alpha=0.51(2)$ for eta meson. 
It is notable that $\alpha_\sigma\sim 0.8$ has a relatively large deviation from the free boson result, $\alpha=1/2$, which may suggest that the sigma meson has a nontrivial dispersion relation even for small momentum. 

Finally, we summarize the masses of the three mesons measured
by the correlation functions: 
\begin{equation}
    \begin{array}{c|c|c|c}
    & \text{pion} & \text{sigma} & \text{eta}\\ \hline
    \text{mass}/g &\,\, 0.431(1) \,\, & \,\, 0.722(6) \,\,&\,\, 0.899(2)
    \end{array}
    \label{tab:mass_cf}
\end{equation}
The numerical results are qualitatively consistent
with the analytic result by bosonization $M_{\pi}<M_{\sigma}<M_{\eta}$.
We also find
\begin{equation}
    M_{\sigma}/M_{\pi}=1.68(2),
\end{equation}
which is close to the prediction
by the sine-Gordon model~Eq.~\eqref{eq:sqrt3-Schwinger}.

\subsection{One-point-function scheme}
\label{subsec:result_1pt}

We consider an alternative way to obtain the mass spectrum without using the two-point correlation functions. 
Let us recall that we are taking the open boundary condition, and we can use those boundaries as the source for excitations from the thermodynamic ground state. 
The boundary effect decays exponentially for the gapped systems, and thus the one-point function of a local operator $\mathcal{O}(x)$ should behave as $\langle \mathcal{O}\rangle+C' e^{-M_{\mathcal{O}} x}$ as the function of the distance $x=an$
from the boundary.
Here, $\langle\mathcal{O}\rangle$ gives the vacuum expectation value in the thermodynamic limit,
and $M_{\mathcal{O}}$ in the exponent gives the lightest particle mass with the same quantum number of $\mathcal{O}(x)$.
In the context of condensed matter physics, it is known that the correlation function
can be obtained from the Fridel oscillation,
which is induced by a boundary effect or a local external field \cite{PhysRevB.54.13495,SHIBATA19971024}.

We note that the $x$-dependence in this method takes the purely exponential form $e^{-M x}$ as the leading behavior for $x\to \infty$. 
This can be easily understood by considering the path integral and the $\pi/2$ rotation of Euclidean spacetime. 
Then, the boundary condition sits at the constant imaginary time and defines the state $\ket{\mathrm{Bdry}}$ with zero momentum. 
Thus, the leading contribution to the imaginary-time correlation function $\bra{\mathrm{Vac}}\mathcal{O}e^{-H|x|}\ket{\mathrm{Bdry}}$
should come from the lightest particle with the zero-momentum projection, giving $e^{-Mx}$. 
This feature has nice compatibility with the low-rank approximation of DMRG. 

\subsubsection{The one-point functions of \texorpdfstring{$\sigma$ and $\eta$}{sigma and eta} at \texorpdfstring{$\theta=0$}{theta=0}}

At $\theta=0$, the boundary condition turns out to be completely invariant under the isospin rotation, and thus the boundary state $\ket{\mathrm{Bdry}}$ does not produce one pion states. 
Therefore, let us here focus on the iso-singlet particles, $\sigma$ and $\eta$, and we will come back to pions later. 

\begin{figure}[tb]
\centering
\includegraphics[scale=0.45]{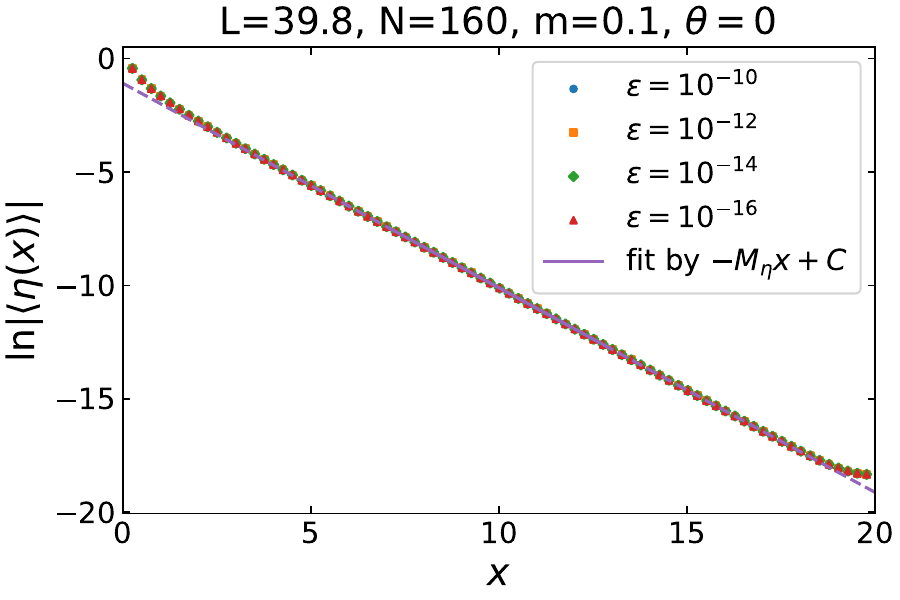}
\caption{\label{fig:1pt_eta}
The one-point function $\ln|\Braket{\eta(x)}|$ of the eta meson
is plotted against $x=an$ with $n=1,\cdots,N/2-1$ for various values of $\varepsilon$.
The number of lattice sites is $N=160$ and the lattice spacing $a$ is determined
so that $L=a(N-1)=39.8$. The result of fitting by $-M_{\eta}x+C$
for $\varepsilon=10^{-16}$ is also plotted by the solid line
inside the range and by the broken line outside.}
\end{figure}

First, we discuss the eta meson as it turns out to be the simplest one. 
Since the $G$-parity is not spontaneously broken, we must have $\langle \eta\rangle=0$ in the thermodynamic limit. 
However, the staggered fermion realizes the $G$-parity (or charge conjugation) as the one-unit lattice translation, and thus the open boundary condition violates the $G$-parity. 
Therefore, the boundary state can be a source of the eta meson, and we evaluate the one-point function $\Braket{\eta(x)}$
of the eta meson operator (\ref{eq:eta_op}) in the range $0<x\le L/2$.
The result is shown in Fig.~\ref{fig:1pt_eta}.
The cutoff parameter is changed from $\varepsilon=10^{-10}$ to $10^{-16}$.
The one-point function decays exponentially with $x$ as expected.
Thus, we fit the data points of $\ln|\Braket{\eta(x)}|$ by $-M_{\eta}x+C$  in the fitting range  $7\leq x\leq13$, 
and the result is 
\begin{equation}
    M_{\eta}=0.9014(1),
\end{equation}
with $C=-1.096(1)$ for the smallest cutoff $\varepsilon=10^{-16}$.
The errors of these values come from the fitting error.
The corresponding fitting curve is shown in Fig.~\ref{fig:1pt_eta} with the purple line.
In this case, we also find that the results for the other values of $\varepsilon$ are consistent within the fitting error. 
Thus, the cutoff dependence does not appear
unlike the case of the correlation functions, and we suspect that this is because MPS can efficiently express purely exponential decay.

\begin{figure}[tb]
\centering
\includegraphics[scale=0.45]{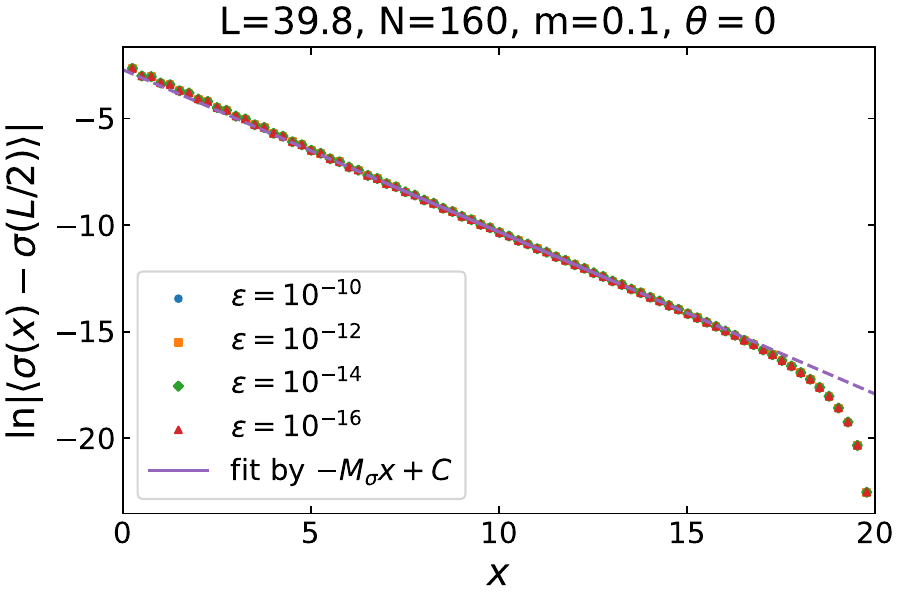}
\caption{\label{fig:1pt_sigma}
The one-point function $\ln|\Braket{\sigma(x)-\sigma(L/2)}|$ of the sigma meson
is plotted against $x=an$ with $n=1,\cdots,N/2-1$ for various values of $\varepsilon$.
The value at $x=L/2$ is subtracted from $\Braket{\sigma(x)}$
to eliminate the constant shift in the bulk.
The number of lattice sites is $N=160$ and the lattice spacing $a$ is determined
so that $L=a(N-1)=39.8$. The result of fitting by $-M_{\sigma}x+C$ is also plotted
by the solid line inside the range and by the broken line outside.}
\end{figure}

Next, we evaluate the one-point function $\Braket{\sigma(x)}$
of the sigma meson  (\ref{eq:sigma_op})  for $0<x\le L/2$. 
We note that $\sigma$ has the same quantum number with the vacuum, and then $\Braket{\sigma(x)}$ is nonzero also in the bulk. 
It behaves as $e^{-Mx+C}+A$ with a constant shift of $A$, 
so we subtract the value $\Braket{\sigma(L/2)}$ at the center $x=L/2$ of the lattice from $\Braket{\sigma(x)}$ to remove the constant.
The result is shown in Fig.~\ref{fig:1pt_sigma}, which indicates the exponential decay as expected.
We fit the data points of $\ln|\Braket{\sigma(x)-\sigma(L/2)}|$ by $-M_{\sigma}x+C$
in the range $7\leq x\leq 13$, and the best-fit parameter is 
\begin{equation}
    M_{\sigma}=0.761(2),
\end{equation}
with $C=-2.71(2)$, which are independent of the value of $\varepsilon$
up to the fitting error.
The result of fitting for $\varepsilon=10^{-16}$ is shown in Fig.~\ref{fig:1pt_sigma} with the purple line.

\subsubsection{The one-point functions of \texorpdfstring{$\pi_3$}{pion} at \texorpdfstring{$\theta=2\pi$}{theta=2pi}}

Let us now come back to the issue of pions. 
As we have argued, the boundary state at $\theta=0$ is neutral under the isospin rotation, and thus it does not produce one-pion states and we have $\langle \pi(x)\rangle =0$ for all $x$. 
Therefore, we need to somehow create the boundary state that transforms nontrivially under the isospin rotation to study pions with the one-point-function scheme.

In this study, we decided to use one of the ground states at $\theta=2\pi$ for this purpose. 
Since the Hamiltonians at $\theta=0$ and $\theta=2\pi$ are unitary equivalent under the periodic boundary condition, the bulk properties are the exactly same between $\theta=0,2\pi$. 
As we have discussed in Section~\ref{subsec:phase_structure}, the ground state at $\theta=2\pi$ is a nontrivial SPT state protected by the isospin $\SU(2)_V/\mathbb{Z}_2$ symmetry,
and thus the boundary states with the open boundary condition have the isospin $1/2$.
This boundary charge can be a source of the pions so that  $\Braket{\pi(x)}$ becomes nonzero. 

About the computational cost, it turns out that the bond dimensions for the MPS are mostly the same at $\theta=0$ and $\theta=2\pi$ when the system size is large enough. 
Therefore, we can obtain the ground state at $\theta=2\pi$ as easily as that of $\theta=0$. 
We, however, observe that the bond dimension at $\theta=2\pi$ increases significantly if the system size is not large enough,
and we suspect its reason is as follows. 
At $\theta=2\pi$, there is $4$-fold degeneracy due to the boundary degrees of freedom,
but they split into the singlet and the triplet states with the energy splitting $\sim e^{-M_\pi L}$. 
That is, the true ground state at finite $L$ has an extra Bell pair between the boundary isospin $1/2$ states, which adds $\ln 2$ to the entanglement entropy. 
When we cut the system at $x=L/2$, this extra $\ln 2$ should be accumulated by the large numbers of small singular values, and thus the bond dimension becomes quite huge just to create the Bell pair between the boundaries. 
If $L$ is large enough, the energy gain by creating the Bell pair becomes negligible, and the DMRG would produce one of the ground states with disentangled boundary states practically.
Thus, the computational cost becomes almost the same as that for the trivial state at $\theta=0$. 

\begin{figure}[tb]
\centering
\includegraphics[scale=0.45]{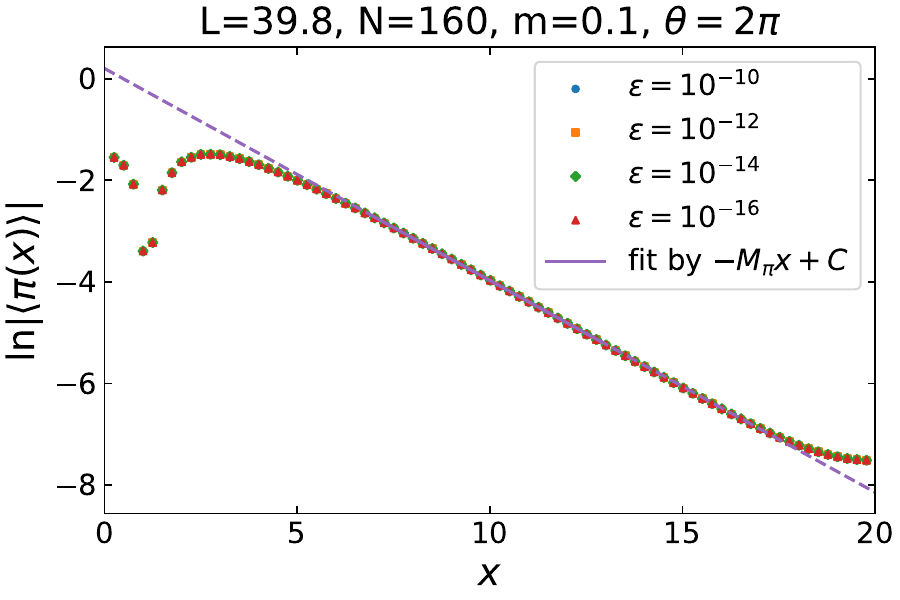}
\caption{\label{fig:1pt_pi}
The one-point function $\ln|\Braket{\pi(x)}|$ of the pion is plotted
against $x=an$ with $n=1,\cdots,N/2-1$ for various values of $\varepsilon$.
The number of lattice sites is $N=160$ and the lattice spacing $a$ is determined
so that $L=a(N-1)=39.8$. We set $\theta=2\pi$ in order to induce the boundary charges,
which make $\Braket{\pi(x)}$ nonzero. The result of fitting by $-M_{\pi}x+C$ is
also plotted by the solid line inside the range and by the broken line outside.}
\end{figure}

Let us now evaluate the one-point function $\Braket{\pi(x)}$ of the pion (\ref{eq:pi_op}) for $0<x\leq L/2$ using the ground state at $\theta=2\pi$.
The result is shown in Fig.~\ref{fig:1pt_pi}.
We again find the exponential decay, and thus fit the data points of
$\ln|\Braket{\pi(x)}|$ by $-M_{\pi}x+C$ in the range $7\leq x\leq 13$.
The result is 
\begin{equation}
    M_{\pi}=0.4175(9),
\end{equation}
with $C=0.203(9)$,
which do not depend on the cutoff $\varepsilon$
up to the fitting error as before.
The fitting result for $\varepsilon=10^{-16}$ is shown in Fig.~\ref{fig:1pt_pi} with the purple line.

Let us summarize the effective masses using the one-point-function scheme:
\begin{equation}
    \begin{array}{c|c|c|c}
    & \text{pion} & \text{sigma} & \text{eta}\\ \hline
    \text{mass}/g &\,\, 0.4175(9) \,\,&\,\, 0.761(2) \,\,&\,\, 0.9014(1)
    \end{array}
    \label{tab:mass_1pt}
\end{equation}
The order of three meson masses is consistent with the analytic prediction. We also find 
\begin{equation}
    M_{\sigma}/M_{\pi}\simeq 1.821(6), 
\end{equation}
which is still close to the WKB prediction, $\sqrt{3}$, with a 5\% deviation.
The significant feature of the method is that the results do not depend on
the cutoff parameter $\varepsilon$ as long as it is sufficiently small. 
Therefore, the systematic error from the cutoff is expected to be small enough.
We do not need to increase the bond dimension so much, unlike the method by the correlation function.

\subsection{Dispersion-relation scheme}
\label{subsec:result_ex}

So far we have studied the mass spectrum by using the local observables
of the ground state, and these methods are applicable both in the path integral and the Hamiltonian formalisms. 
As the third method for computing the mass spectrum, let us take a different approach that is specific to the Hamiltonian formalism: 
We compute the excited states as explained in Section~\ref{sec:method}, and then determine the mass spectrum from the dispersion relation.

The low-lying excited states correspond to one-particle excitations.
For example, the zero mode of the lightest meson, namely the pion, is expected to be obtained as the first excited state. 
We can also obtain the states with nonzero momentum $K$, and we can fit the data with the dispersion relation $\Delta E\simeq\sqrt{K^{2}+M_{\pi}^{2}}$ to obtain the pion mass.
As we go to the higher excited states, we will encounter one-particle states of the sigma and eta mesons. 
They can be distinguished by measuring quantum numbers,
such as the isospin and $G$-parity. Thus, we can compute the mass spectrum from the dispersion relation by generating the excited states. 

We note that our computation is done in the finite open interval, and thus the momentum is not a good quantum number. 
Also, there may exist a nontrivial contribution to the excitation energy from the boundaries. We are neglecting those subtleties in this work, but, surprisingly, it turns out that the numerical results are almost consistent with those with the previous two methods. 

\begin{figure}[tb]
\centering
\includegraphics[scale=0.45]{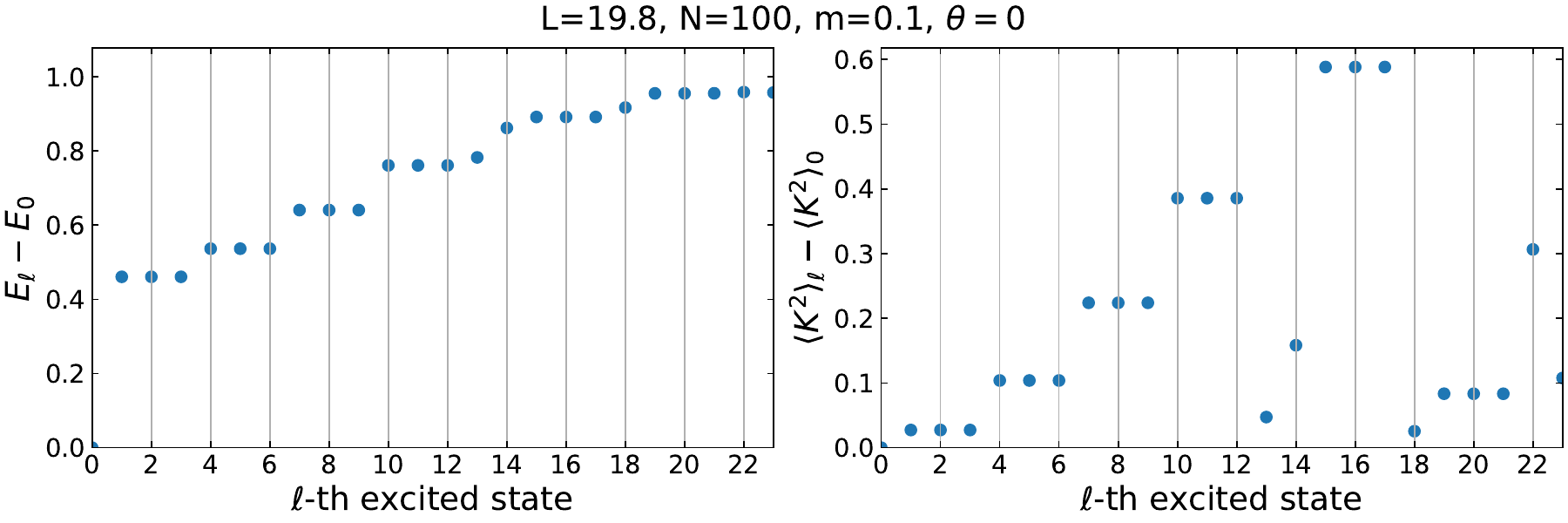}
\caption{\label{fig:E_and_K2}
(Left) The energy gap $\Delta E_{\ell}=E_{\ell}-E_{0}$
is plotted against the level of the excited state $\ell$.
(Right) The square of total momentum
$\Delta K_{\ell}^2=\Braket{K^2}_{\ell}-\Braket{K^{2}}_{0}$
is plotted against $\ell$ after subtracting the result for the ground state.}
\end{figure}

We generated the MPS up to the 23rd excited state at $\theta=0$ with the small physical volume $L=19.8$.
The energy gap $\Delta E_{\ell}=E_{\ell}-E_{0}$ of the $\ell$-th excited state is shown
in the left panel of Fig.~\ref{fig:E_and_K2}. We also measured the square of total momentum $K^{2}$ defined by (\ref{eq:K_staggered}).
We note that its ground-state expectation value $\Braket{K^{2}}_{0}\simeq 0.46$ is nonzero because of the boundary effect and maybe also due to lattice artifacts, and thus we subtract $\Braket{K^{2}}_{0}$ from $\Braket{K^{2}}_{\ell}$ of the excited states. 
The result is plotted in the right panel of Fig.~\ref{fig:E_and_K2}.
From these results, we find many triply degenerated states,
which are candidates for the states of the pion.
There are a few singlet states as well, which are candidates for the eta and sigma mesons. 

To identify the states, we measure the expectation values of the isospin operators, $\bm{J}^{2}$ and $J_{z}$, the parity $P$ and the $G$-parity $G=Ce^{i\pi J_{y}}$ defined in Section~\ref{subsec:global_obs}.
We note that the DMRG does not produce the states in a diagonal basis for these quantities. 
We diagonalize the $3\times3$ matrix
$\Braket{\psi_{\ell_1}|J_{z}|\psi_{\ell_2}}$ in each triplet to compute expectation values in the $J_z$ basis.\footnote{In computing the expectation value of the $G$-parity, we find it easier to do it in the $J_y$ basis instead of the $J_z$ basis because $G=Ce^{i\pi J_y}$, and we thus performed it in the $J_y$ basis.}\textsuperscript{,}\footnote{It is possible that triplets and singlets are also mixed in the DMRG if their energies are close. In fact, the states for $\ell=19,\cdots,23$
are mostly degenerated. We separated one triplet and two singlets out of them
by diagonalization of $\Braket{\psi_{\ell_1}|\bm{J}^{2}|\psi_{\ell_2}}$ and of $\Braket{\psi_{\ell_1}|C|\psi_{\ell_2}}$.}

\begin{table}[tb]
\centering
\begin{tabular}{|c|c|c|c|c|}\hline $\ell$ & $\bm{J}^2$ & $J_z$ & $G$ & $P$ \tabularnewline\hline \hline 1 &  2.00000004 &  0.99999997 &  0.27872443 & -6.819$\times {10}^{-8}$ \tabularnewline2 &  2.00000012 & -0.00000000 &  0.27872416 & -6.819$\times {10}^{-8}$ \tabularnewline3 &  2.00000004 & -0.99999996 &  0.27872443 & -6.819$\times {10}^{-8}$ \tabularnewline\hline 4 &  2.00000007 &  0.99999999 &  0.27736066 &  7.850$\times {10}^{-8}$ \tabularnewline5 &  2.00000006 &  0.00000000 &  0.27736104 &  7.850$\times {10}^{-8}$ \tabularnewline6 &  2.00000009 & -0.99999998 &  0.27736066 &  7.850$\times {10}^{-8}$ \tabularnewline\hline 7 &  2.00000010 &  1.00000000 &  0.27536687 & -8.838$\times {10}^{-8}$ \tabularnewline8 &  2.00000002 &  0.00000000 &  0.27536702 & -8.837$\times {10}^{-8}$ \tabularnewline9 &  2.00000007 & -0.99999998 &  0.27536687 & -8.838$\times {10}^{-8}$ \tabularnewline\hline 10 &  2.00000007 &  0.99999998 &  0.27356274 &  9.856$\times {10}^{-8}$ \tabularnewline11 &  2.00000005 &  0.00000001 &  0.27356277 &  9.856$\times {10}^{-8}$ \tabularnewline12 &  2.00000007 & -0.99999999 &  0.27356274 &  9.856$\times {10}^{-8}$ \tabularnewline\hline 15 &  1.99999942 &  0.99999966 &  0.27173470 & -1.077$\times {10}^{-7}$ \tabularnewline16 &  2.00000052 &  0.00000000 &  0.27173482 & -1.077$\times {10}^{-7}$ \tabularnewline17 &  2.00000015 & -1.00000003 &  0.27173470 & -1.077$\times {10}^{-7}$ \tabularnewline\hline 19 &  2.00009067 &  1.00004377 &  0.27717104 & -3.022$\times {10}^{-8}$ \tabularnewline20 &  2.00002578 & -0.00000004 &  0.27717020 & -3.023$\times {10}^{-8}$ \tabularnewline21 &  2.00003465 & -1.00001622 &  0.27717104 & -3.023$\times {10}^{-8}$ \tabularnewline\hline \end{tabular}\vspace{-1.3em}
\caption{\label{tab:triplet}
The quantum numbers of the isospin triplet states.
The index $\ell$ comes from the level of each state in the original
basis. The rows of the table are separated into each triplet.}
\end{table}
\begin{table}[tb]
\centering
\begin{tabular}{|c|c|c|c|c|}\hline $\ell$ & $\bm{J}^2$ & $J_z$ & $G$ & $P$ \tabularnewline\hline \hline 0 &  0.00000003 & -0.00000000 &  0.27984227 &  3.896$\times {10}^{-7}$ \tabularnewline\hline 13 &  0.00000003 &  0.00000000 &  0.27865844 &  1.273$\times {10}^{-7}$ \tabularnewline\hline 14 &  0.00000003 &  0.00000000 &  0.27508176 & -2.765$\times {10}^{-8}$ \tabularnewline\hline 18 &  0.00000028 &  0.00000006 & -0.27390909 & -6.372$\times {10}^{-7}$ \tabularnewline\hline 22 &  0.00001537 &  0.00000115 &  0.26678987 &  7.990$\times {10}^{-8}$ \tabularnewline\hline 23 &  0.00003607 & -0.00000482 & -0.27664779 &  5.715$\times {10}^{-7}$ \tabularnewline\hline \end{tabular}\vspace{-1.3em}
\caption{\label{tab:singlet}
The quantum numbers of the isospin singlet states.}
\end{table}

The expectation values of $\bm{J}^{2}$, $J_{z}$, $G$, and $P$ in the $J_{z}$ basis are listed in Tables~\ref{tab:triplet} and~\ref{tab:singlet}
for iso-triplets and iso-singlets, respectively.
The index $\ell$ comes from the level of the state on the original random basis. We find that $|G|\neq1$ because of $|C|\neq1$ by the effect of the boundary. 
Hopefully, the sign of $G$ can be assumed to remember the original quantum number~\cite{Banuls:2013jaa},
and, if it is true, we can still identify the $G$-parity. 
This point will be discussed more in details in Appendix~\ref{sec:C_test}. 
We identify the lowest triplet $\ell=1,2,3$ as the lowest modes of
the pions ($\pi^{+}$, $\pi^{0}$, $\pi^{-}$) since they have the quantum numbers
consistent with the pion, namely $J^{PG}=1^{-+}$ and $J_{z}=0,\pm1$.
For the iso-singlets shown in Table \ref{tab:singlet}, 
we find that the $\ell=13$ state has the quantum number consistent with the sigma meson,
namely $J^{PG}=0^{++}$ and $J_{z}=0$.
The $\ell=18$ state is consistent with the eta meson
with $J^{PG}=0^{--}$ and $J_{z}=0$.
We identify these singlets with the lowest modes of the sigma and eta mesons.

\begin{figure}[t]
\centering
\includegraphics[scale=0.45]{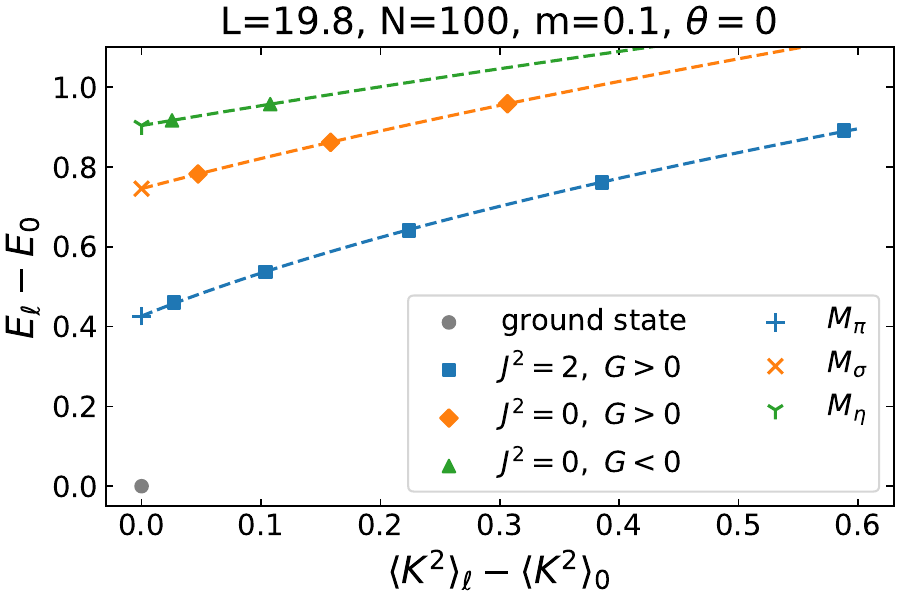}
\caption{\label{fig:E_vs_K2}
The energy gap $\Delta E_{\ell}$ is plotted against
the square of total momentum $\Delta K_{\ell}^{2}$. The states with
the same isospin and $G$-parity are plotted by the same symbol. Then
each state is identified with the pion, sigma, or eta meson. We fit the data
for each meson by $\Delta E=\sqrt{b^2 \Delta K^{2}+M^{2}}$. The results
are shown by the broken lines. The values of $M$ for each meson are
also plotted as the endpoints of the fitting lines.}
\end{figure}

After identifying the quantum numbers, we plot the energy gap
$\Delta E_{\ell}=E_{\ell}-E_0$ against the momentum square
$\Delta K_{\ell}^{2}=\Braket{K^2}_{\ell}-\Braket{K^2}_0$
to obtain the dispersion relation as shown in Fig.~\ref{fig:E_vs_K2}.
The states with the same isospin $\bm{J}^{2}$ and $G$-parity are
plotted by the same symbol.\footnote{
The triplet for $\ell=19,20,21$ is not shown in this plot since it is
not of the state of the single pion. We expect that the triplet comes from
the pion scattering state, which was discussed in \cite{Harada:1993va}.}
Then we fit the data points by
$\Delta E=\sqrt{b^2 \Delta K^{2}+M^{2}}$ with fitting parameters $M$ and $b$.
The fitting result of $M$ can be regarded as the mass of the corresponding meson
as an extrapolation to $\Delta K^{2}\rightarrow 0$.
We obtained $M_{\pi}=0.426(2)$, $b_{\pi}=1.017(4)$ for the pion;
and $M_{\sigma}=0.7456(5)$, $b_{\sigma}=1.087(2)$ for the sigma meson
with the fitting error.
The fitting for the eta meson is simply solving an equation
since there are only two data points.
The result are $M_{\eta}=0.904$ and $b_{\eta}=0.962$.
We summarize the masses of the mesons determined by the energy gap
of the excited states:
\begin{equation}
    \begin{array}{c|c|c|c}
    & \text{pion} & \text{sigma} & \text{eta}\\ \hline
    \text{mass}/g &\,\, 0.426(2) \,\,&\,\, 0.7456(5) \,\,&\,\, 0.904
    \end{array}
    \label{tab:mass_ex}
\end{equation}
We find the mass ratio 
\begin{equation}
    M_{\sigma}/M_{\pi}\simeq 1.75(1)
\end{equation} 
from this result, which is close to the WKB prediction $\sqrt{3}$.

\section{Conclusion and Discussion}
\label{sec:summary}

In this paper, we work on three independent methods to compute the mass spectrum of lattice gauge theories in the Hamiltonian formalism, 
which apply to tensor networks and quantum computation.
The methods are tested in the massive 2-flavor Schwinger model at $\theta=0$, some of which properties are analogous to the ones of $4$d QCD.
The two species of fermion play roles of up and down quarks,
and the composite particles (mesons) appear as triplets or singlets of the $\SU(2)_V/\mathbb{Z}_2$ isospin symmetry.
We used the tensor network, in particular, DMRG for numerical simulation.

\begin{figure}[tb]
\centering
\includegraphics[scale=0.45]{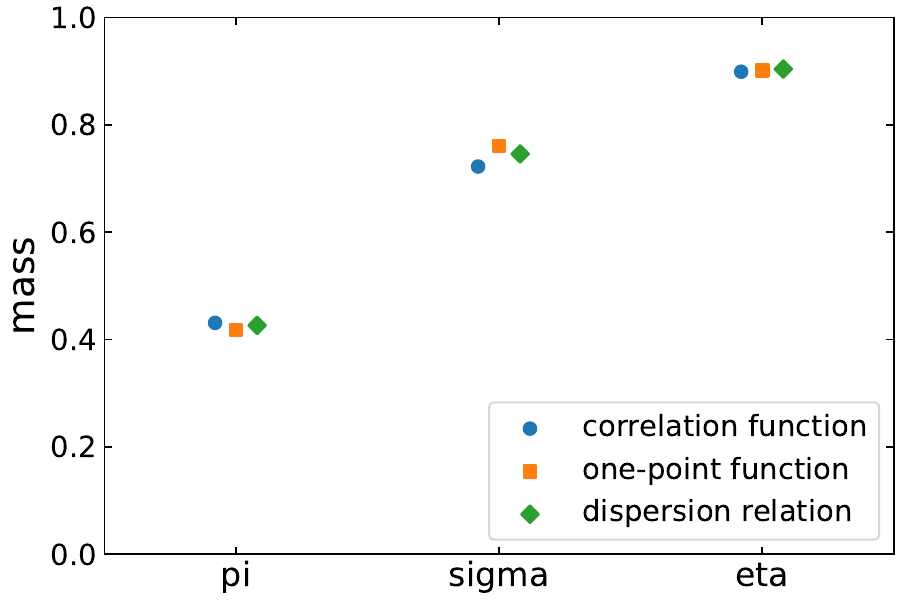}
\caption{\label{fig:mass_comp}
The masses of the pion, sigma, and eta meson obtained by the three independent methods are compared.
Each result is obtained with the given finite lattice spacing.
We also put the error bar of the fitting error for the correlation-function scheme, but it is too small to be seen.}
\end{figure}

We obtained the masses of the pion, sigma, and eta meson by the three methods, 
and the results are summarized in Fig.~\ref{fig:mass_comp}.
We find that the results are roughly consistent with each other taking into account possible systematic errors for each method, such as the continuum and infinite-volume limits.
In addition, all the results show the relation $M_{\pi}<M_{\sigma}<M_{\eta}$, which agrees with the analytic prediction by the bosonization technique.
The order of the eta meson mass $M_{\eta}\sim 0.9$ is consistent with $M_{\eta}\sim\mu$ since $\mu\sim 0.8$ in the current setup.
We also find that the relation between the masses of the pion and sigma mesons
is $M_\sigma/M_\pi=1.68(2),1.821(6),1.75(1)$
by the correlation-function scheme, the one-point-function scheme,
and the dispersion-relation scheme, respectively. 
We note that the errors in the above values only contain the fitting error, and there should be further systematic errors potentially coming from the finite lattice spacing, the finite-volume effect, the open boundary condition, the cutoff of the bond dimension, etc. 
These results are close to the WKB-based formula \eqref{eq:sqrt3-Schwinger}, $M_{\sigma}/M_{\pi}=\sqrt{3}$, within not more than a 5\% deviation.
It is, honestly, very surprising that the semiclassical analysis gives the almost correct answer outside the range of its validity, and it would be theoretically interesting to uncover the reason behind its success. 

Let us discuss the advantages and difficulties of each method and the potential applications to other models.
The first one, the correlation-function scheme, is the straightforward generalization of the technique in Lagrangian formalism.
The advantage of this method is a wide range of applicability to various models.
We can obtain the meson masses from correlation functions on a lattice with any dimensions, volume, and boundary condition.
Furthermore, the correlation function accepts the off-diagonal element such as $\Braket{\mathcal{O}(x)\mathcal{O}^{\prime}(y)}$.
This feature will be useful when we turn on $\theta\neq 0$ in the 2-flavor Schwinger model.
The reason is that the meson operators become nontrivial mixtures of $S_f(n)$ and $PS_f(n)$ depending on $\theta$.
In this case, we need to measure the correlation matrix of the operators and diagonalize it to extract the mode of each meson.
However, our numerical results suggest that the bond dimension of MPS has to be sufficiently large to reproduce the correct asymptotic behavior of the correlation function.
In particular, the computational cost increases rapidly as the system approaches a gapless phase, for example, $m\sim 0$ or $\theta\sim\pi$.
Thus, the tensor network (MPS) is not an efficient approach to computing the mass spectrum by using correlation functions.\footnote{
It is possible that other types of tensor networks, such as MERA, may be still useful in this method.}
On the other hand, an ideal quantum computer is free from such a restriction of the bound dimension. 
Thus, the correlation function may be the first option in the era of practical quantum computation of field theories in this sense, though to avoid the finite volume effect for the two-point function we need a sizable scale computer.

The second method, the one-point-function scheme, makes good use of the boundary effect rather than eliminating it.
The results turn out to be insensitive to the bond dimension, and thus
we have to increase neither the lattice size nor the bond dimension so much.
Furthermore, the evaluation of the local one-point function is generally easier than that of the long-range correlation function.
Thus, this is the most economical one among the three methods.
We note, however, that we have to prepare suitable boundary conditions, such as defects, impurities, or external fields, to compute the mass spectrum with this one-point-function scheme, which requires good physical insights for the system of interest. 
In our case, the open boundary at $\theta=0$ can be regarded as a source of the iso-singlet mesons, $\sigma$, and $\eta$,
but we have to set $\theta=2\pi$ to induce the boundary excitation as a source of the iso-triplet mesons, $\pi_{a}$.
We should also note that we cannot obtain information on the off-diagonal correlators in the one-point-function scheme. 
When $\theta=0$, the off-diagonal correlators are unimportant because $\pi$, $\sigma$, and $\eta$ have different quantum numbers, but they should become important at generic values of $\theta$ because the $G$-parity is no longer a good quantum number. 

The third method, the dispersion-relation scheme, is the distinctive strategy of Hamiltonian formalism.
We can obtain various states heuristically without knowing what kind of mesons appear in the spectrum.
Once we generate the excited states, it is straightforward to measure various observables such as energy, momentum, and quantum numbers.
The states are identified by using these pieces of information.
Furthermore, we can investigate the wave function to distinguish the $s$-wave or $p$-wave states.
In this method, however, it is difficult to increase the system size or the spatial dimensions.
The reason is that we have to generate an increasing number of states to search for different mesons.
For example, in our setup, we encounter the $3\times 4$ states of the pion before obtaining the sigma meson at $\ell=13$.
The momentum $K$ is discretized as $K\sim 2\pi\kappa/L$ for $\kappa=1,2,\cdots$ in the finite system with the size $L$.
If $L$ is increased, the number of pion states in a certain range of energy grows up.
Thus, we have to generate more excited states to reach the state of the sigma meson.
As for higher dimensions, there are momentum excitations in each spatial direction, which result in an additional degeneracy.
We expect that there is a way to avoid this issue by modifying the strategy.
For example, if we are interested in a specific meson, it is more effective to generate excited states with a constraint on the quantum number to skip mesons out of interest.

In this work, we have computed the mass spectrum at $\theta = 0$. 
We note that we have neglected many systematic errors, and thus there is plenty of room for improvement. 
As a physics, extending our investigation to $\theta \neq 0$ should be interesting,
where the sign problem arises in naive applications of Monte Carlo simulations.
The presence of $\theta$ introduces some differences compared to the $\theta = 0$ case.
Firstly, the mass of the pion, which corresponds to the gap of the system, decreases as $\theta\to \pi$.
Consequently, we may need to increase the bound dimension of MPS, leading to higher computational costs.
Secondly, the parity and $G$-parity are no longer good quantum numbers for $\theta \neq 0$, and the scalar and pseudo-scalar operators have a nontrivial mixture.
To handle this situation, we should measure the correlation matrix between these operators and diagonalize it. 
Although distinguishing the excited states, especially $\sigma$ and $\eta$, seems to become tricky,
exploring the changes in the spectrum promises intriguing insights.
Despite these subtleties, we expect that it is still possible to apply the three methods
to compute the mass spectrum including the theta term, and the results at $\theta \neq 0$ will be reported elsewhere.
Needless to say, it is very desirable that future developments of these techniques eventually enable us to compute the hadron spectrum of $4$d strongly-coupled gauge theories having the sign problem in the conventional Monte Carlo methods.

\acknowledgments
We would like to thank S.~Aoki, M.~Honda, T.~Nishino, and K.~Okunishi for their useful discussions.
The numerical calculations were carried out on
XC40 at YITP in Kyoto University
and the PC clusters at RIKEN iTHEMS.
The work of A.~M. is supported by FY2022 Incentive Research Projects of RIKEN.
The work of E.~I. is supported by JST PRESTO Grant Number JPMJPR2113, 
JST Grant Number JPMJPF2221, %COI-NEXT
JSPS KAKENHI (S) Grant number 23H05439, %Kiban S
JSPS Grant-in-Aid for Transformative Research Areas (A) JP21H05190, and 
Program for Promoting Researches on the Supercomputer Fugaku” (Simulation for basic science: approaching the new quantum era) Grant number JPMXP1020230411.
The work of Y.~T. is supported by JSPS KAKENHI Grant number, 22H01218.
This work is supported by Center for Gravitational Physics and Quantum Information (CGPQI) at YITP.

\appendix

\section{Operators in the spin representation}
\label{sec:Spin_op}

In this appendix, we show the spin representations of the Hamiltonian
and operators defined in Section~\ref{sec:formulation}
after the Jordan-Winger transformation \eqref{eq:JW_1} and \eqref{eq:JW_2}.
For later convenience, we first show the transformation of some local operators,
\begin{equation}
\chi_{f,n}^{\dagger}\chi_{f,n}=\sigma_{f,n}^{+}\sigma_{f,n}^{-}=\frac{\sigma_{f,n}^{z}+1}{2},
\end{equation}
\begin{equation}
\chi_{1,n}^{\dagger}\chi_{1,n+1}-\chi_{1,n+1}^{\dagger}\chi_{1,n}=\sigma_{1,n}^{+}\sigma_{2,n}^{z}\sigma_{1,n+1}^{-}-\sigma_{1,n}^{-}\sigma_{2,n}^{z}\sigma_{1,n+1}^{+},
\end{equation}
\begin{equation}
\chi_{2,n}^{\dagger}\chi_{2,n+1}-\chi_{2,n+1}^{\dagger}\chi_{2,n}=\sigma_{2,n}^{+}\sigma_{1,n+1}^{z}\sigma_{2,n+1}^{-}-\sigma_{2,n}^{-}\sigma_{1,n+1}^{z}\sigma_{2,n+1}^{+}.
\end{equation}
The product of $\sigma^{z}$ in the Jordan-Winger transformation mostly cancels
each other in the fermion bilinears.
Let us start with the Hamiltonian.
Using the relation above, the gauge part $H_{J}$ \eqref{eq:H_J} is transformed as
\begin{equation}
H_J =\frac{J}{4}\sum_{n=0}^{N-2}
\left[\sum_{f=1}^{N_{f}}\sum_{k=0}^{n}\sigma_{f,k}^{z}
+N_{f}\frac{(-1)^{n}+1}{2}+\frac{\theta}{\pi}\right]^2.
\end{equation}
The the fermion kinetic term $H_{w}$ \eqref{eq:H_w} and the mass term $H_{m}$ \eqref{eq:H_m}
are given by
\begin{align}
H_{w} & =-iw\sum_{n=0}^{N-2}\left(\sigma_{1,n}^{+}\sigma_{2,n}^{z}\sigma_{1,n+1}^{-}-\sigma_{1,n}^{-}\sigma_{2,n}^{z}\sigma_{1,n+1}^{+}\right.\nonumber \\
 & \hphantom{=-iw\sum_{n=0}^{N-2}}\left.+\sigma_{2,n}^{+}\sigma_{1,n+1}^{z}\sigma_{2,n+1}^{-}-\sigma_{2,n}^{-}\sigma_{1,n+1}^{z}\sigma_{2,n+1}^{+}\right),
\end{align}
\begin{equation}
H_{m} =\frac{m_{\mathrm{lat}}}{2}\sum_{f=1}^{N_{f}}\sum_{n=0}^{N-1}(-1)^{n}\sigma_{f,n}^{z}+\frac{m_{\mathrm{lat}}}{2}N_{f}\frac{1-(-1)^{N}}{2}.
\end{equation}
Then the total Hamiltonian is a sum of them,
\begin{equation}
H=H_{J}+H_{w}+H_{m}.\label{eq:H_spin}
\end{equation}
For practical use, $H_{J}$ can be decomposed into the quadratic term
of $\sigma^{z}$, the linear term of $\sigma^{z}$, and the constant
term by expanding the square. They can be summarized as follows:
\begin{equation}
H_{J}=H_{J}^{(2)}+H_{J}^{(1)}+H_{J}^{(0)},
\end{equation}
\begin{equation}
H_{J}^{(2)}=\frac{J}{2}\sum_{f=1}^{N_{f}}\sum_{j=0}^{N-3}\sum_{k=j+1}^{N-2}(N-k-1)\sigma_{f,j}^{z}\sigma_{f,k}^{z}+\frac{J}{4}\sum_{f\neq f^{\prime}}\sum_{n=0}^{N-2}\sum_{j,k=0}^{n}\sigma_{f,j}^{z}\sigma_{f^{\prime},k}^{z},
\end{equation}
\begin{equation}
H_{J}^{(1)}=\frac{J}{2}\sum_{f=1}^{N_{f}}\sum_{k=0}^{N-2}\left[\left(\frac{N_{f}}{2}+\frac{\theta}{\pi}\right)(N-k-1)+\frac{N_{f}}{2}\frac{(-1)^{N}+(-1)^{k}}{2}\right]\sigma_{f,k}^{z},
\end{equation}
\begin{align}
H_{J}^{(0)} & =\frac{JN_{f}}{4}\frac{N(N-1)}{2} \nonumber \\
 & +\frac{JN_{f}}{2}\left(\frac{N_{f}}{4}+\frac{\theta}{2\pi}\right)\left[\frac{(-1)^{N}-1}{2}+N\right]+J\left(\frac{\theta}{2\pi}\right)^{2}(N-1).
\end{align}
The spin Hamiltonian contains the non-local interactions which come from the Gauss law.
It is not obvious whether the ground state can be described efficiently by MPS.

Next, we map the observables by the Jordan-Winger transformation.
The local scalar condensate (\ref{eq:S_staggered})
and the pseudo-scalar condensate (\ref{eq:PS_staggered}) are transformed as
\begin{equation}
S_{f}(n)=\frac{1}{8a}(-1)^{n}(-\sigma_{f,n-1}^{z}+2\sigma_{f,n}^{z}-\sigma_{f,n+1}^{z}),
\end{equation}
\begin{align}
PS_{1}(n)=\frac{i}{4a}(-1)^{n} & \left(\sigma_{1,n-1}^{+}\sigma_{2,n-1}^{z}\sigma_{1,n}^{-}-\sigma_{1,n-1}^{-}\sigma_{2,n-1}^{z}\sigma_{1,n}^{+}\right.\nonumber \\
 & \left.-\sigma_{1,n}^{+}\sigma_{2,n}^{z}\sigma_{1,n+1}^{-}+\sigma_{1,n}^{-}\sigma_{2,n}^{z}\sigma_{1,n+1}^{+}\right),
\end{align}
\begin{align}
PS_{2}(n)=\frac{i}{4a}(-1)^{n} & \left(\sigma_{2,n-1}^{+}\sigma_{1,n}^{z}\sigma_{2,n}^{-}-\sigma_{2,n-1}^{-}\sigma_{1,n}^{z}\sigma_{2,n}^{+}\right.\nonumber \\
 & \left.-\sigma_{2,n}^{+}\sigma_{1,n+1}^{z}\sigma_{2,n+1}^{-}+\sigma_{2,n}^{-}\sigma_{1,n+1}^{z}\sigma_{2,n+1}^{+}\right).
\end{align}
We can also map the isospin operators
\eqref{eq:J_z}, \eqref{eq:J_plus}, and \eqref{eq:J_minus} as follows:
\begin{equation}
J_{z}=\frac{1}{4}\sum_{n=0}^{N-1}(\sigma_{1,n}^{z}-\sigma_{2,n}^{z}),
\end{equation}
\begin{equation}
J_{+}=i\sum_{n=0}^{N-1}\sigma_{1,n}^{+}\sigma_{2,n}^{-},
\end{equation}
\begin{equation}
J_{-}=-i\sum_{n=0}^{N-1}\sigma_{2,n}^{+}\sigma_{1,n}^{-}.
\end{equation}

Finally, we consider the Jordan-Winger transformation of the total momentum operator
(\ref{eq:K_staggered}). Each term in the sum is transformed as follows:
\begin{align}
& \chi_{1,n-1}^{\dagger}\chi_{1,n+1}-\chi_{1,n+1}^{\dagger}\chi_{1,n-1} \nonumber \\
& = -\sigma_{1,n-1}^{+}\sigma_{2,n-1}^{z}\sigma_{1,n}^{z}\sigma_{2,n}^{z}\sigma_{1,n+1}^{-}
  +\sigma_{1,n-1}^{-}\sigma_{2,n-1}^{z}\sigma_{1,n}^{z}\sigma_{2,n}^{z}\sigma_{1,n+1}^{+},
\end{align}
\begin{align}
& \chi_{2,n-1}^{\dagger}\chi_{2,n+1}-\chi_{2,n+1}^{\dagger}\chi_{2,n-1} \nonumber \\
& = -\sigma_{2,n-1}^{+}\sigma_{1,n}^{z}\sigma_{2,n}^{z}\sigma_{1,n+1}^{z}\sigma_{2,n+1}^{-}
  +\sigma_{2,n-1}^{-}\sigma_{1,n}^{z}\sigma_{2,n}^{z}\sigma_{1,n+1}^{z}\sigma_{2,n+1}^{+}.
\end{align}
Thus, the total momentum is given by the combination of five Pauli matrices,
\begin{align}
K=\frac{i}{4a}\sum_{n=1}^{N-2} & \left(\sigma_{1,n-1}^{-}\sigma_{2,n-1}^{z}\sigma_{1,n}^{z}\sigma_{2,n}^{z}\sigma_{1,n+1}^{+}-\sigma_{1,n-1}^{+}\sigma_{2,n-1}^{z}\sigma_{1,n}^{z}\sigma_{2,n}^{z}\sigma_{1,n+1}^{-}\right.\nonumber \\
 & \left.+\sigma_{2,n-1}^{-}\sigma_{1,n}^{z}\sigma_{2,n}^{z}\sigma_{1,n+1}^{z}\sigma_{2,n+1}^{+}-\sigma_{2,n-1}^{+}\sigma_{1,n}^{z}\sigma_{2,n}^{z}\sigma_{1,n+1}^{z}\sigma_{2,n+1}^{-}\right).
\end{align}
It is possible to construct MPOs systematically from these spin representations.

\section{Charge conjugation operator in the 1-flavor Schwinger model}
\label{sec:C_test}

The charge conjugation operator $C$ defined by (\ref{eq:C_spin})
does not commute with the Hamiltonian under the open boundary condition. 
This is because the charge conjugation for the staggered fermion must incorporate the one-unit lattice translation, and thus it is not an on-site symmetry in our regularization scheme. 
As a consequence, the expectation value of $C$ does not become $\pm1$. 
However, this is an important quantum number to diagnose the type of mesons, and we have assumed in
Section~\ref{subsec:result_ex} that we can diagnose the quantum number by the sign of $\langle C\rangle$. 
Although we have no theoretical justifications for this prescription, let us test it in the $1$-flavor Schwinger model to give some evidence for its reasonableness. 

\begin{figure}[tb]
\centering
\includegraphics[scale=0.45]{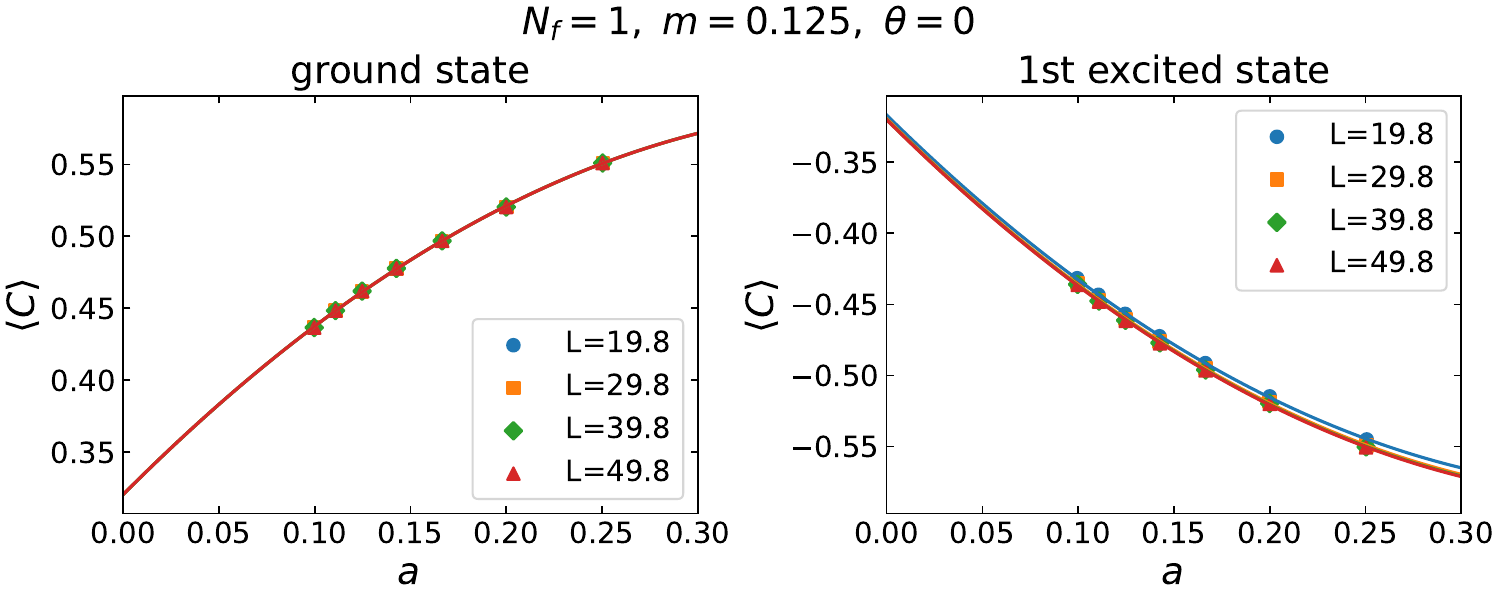}
\caption{\label{fig:C_nf1_cont}
The expectation values of the charge conjugation $C$ for the ground state (left)
and the first excited state (right) are plotted against the lattice spacing $a$.
The number of lattice sites $N$ is chosen to fix the physical length $L=(N-1)a$.
Each symbol corresponds to a different value of $L$.
We set $\theta=0$ and $m=0.125$ in this analysis.
The fitting results by the quadratic are also shown by the solid lines.}
\end{figure}

First, we investigate the behavior of $C$ in the continuum limit.
We generated the MPS of the ground state and the 1st excited state
of the 1-flavor Schwinger model at $\theta=0$ by DMRG.
The lattice spacing $a$ is changed around $0.1\lesssim a\lesssim0.25$.
The number of lattice sites $N$ is chosen to fix the physical system size $L=(N-1)a$.
We compute the expectation values of $C$ for these MPS.
The results are shown in Fig.~\ref{fig:C_nf1_cont}.
The different symbols correspond to the results for different $L$ in the plots.
We fitted the data points for each $L$ by the quadratic function
$f(a)=c_{0}+c_{1}a+c_{2}a^{2}$. The fitting results are also plotted in
Fig.~\ref{fig:C_nf1_cont} by the solid lines. For $L=49.8$, we obtained
the continuum limit $\Braket{C}_{a\rightarrow0}=0.321(3)$
for the ground state and $-0.320(3)$ for the 1st excited state.
The results with the other $L$ agree with these values within the error.
Thus, we confirmed that the expectation value of $C$ is a finite value
in the continuum limit and is not sensitive to $L$.

\begin{figure}[tb]
\centering
\includegraphics[scale=0.45]{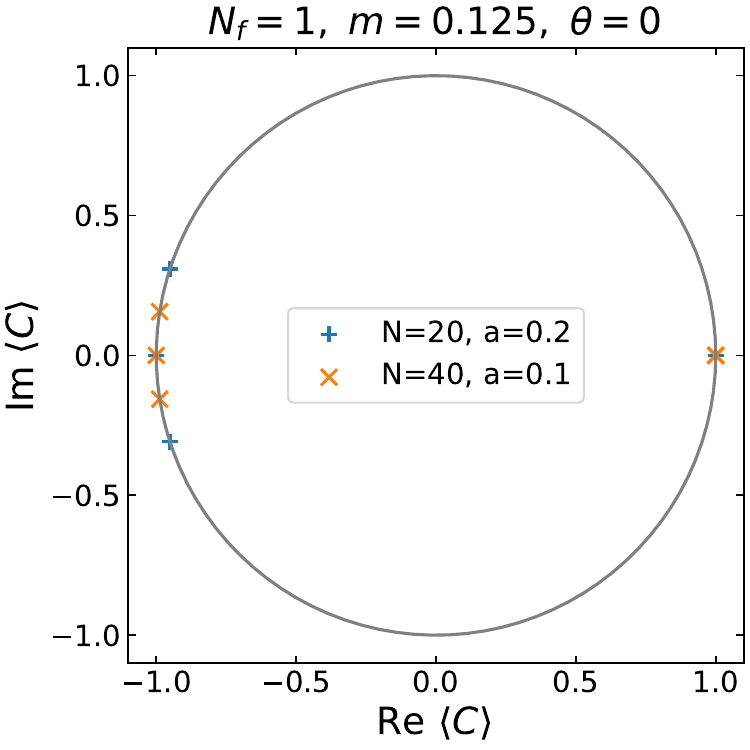}
\caption{\label{fig:C_nf1_pbc}
The expectation values of $C$ are plotted
on the complex plane. The different symbols represent the results
for the different values of the spacing $a$.
All the data points turned out to be on the unit circle.}
\end{figure}

Next, let us discuss the effect of the boundary on $C$.
We consider a further simplified model, the free fermion on the periodic lattice.
The model is obtained from the 1-flavor Schwinger model
with the periodic boundary condition by setting $g=0$
and adding the hopping term between $n=0$ and $n=N-1$ site.
In fact, it is hard to adopt the p.b.c. in the current DMRG method
due to the artificial long-range interaction between both ends of MPS.
Thus, we choose small sizes of the lattice $N=20$ and $40$ for this analysis.
The corresponding lattice spacings are $a=0.2$ and $0.1$
for the fixed physical length $L=Na=4$.
We generate the ground state and the excited states up to the level $\ell=4$.
The four excited states turned out to be degenerated.
Thus, we compute $\Braket{C}_{\ell,\ell^{\prime}}$ including the off-diagonal elements,
and diagonalize the result as the $4\times4$ matrix.
The eigenvalues are shown in Fig.~\ref{fig:C_nf1_pbc}.
We found that $\Braket{C}=1$ for the ground state
and $\Braket{C}=\pm1,\alpha\pm i\beta$ for the excited states.
These complex values satisfy $|\alpha|^{2}+|\beta|^{2}=1$ as we can see in the plot.
The imaginary part $\beta$ becomes smaller as $a$ is decreased,
which suggests that we will obtain $\Braket{C}\rightarrow\pm1$ in the continuum limit.
 
\section{Arrangement of flavors on MPS}
\label{sec:order_of_index}

In the spin representation of the Hamiltonian \eqref{eq:H_spin}
of the 2-flavor Schwinger model,
each spin has the site index $n$ and the flavor index $f$. To apply DMRG,
we arrange these spins on the 1d lattice with the single index $(f,n)\rightarrow i$.
Although the ordering of the indices does not affect the physics,
it can affect the necessary bound dimensions, and thus calculation cost depends on it.
In this work, we assign the index $i$ to $(f,n)$ as
\begin{equation}
i=N_{f}n+f-1=0,1,\cdots,N_{f}N,\label{eq:staggered_order}
\end{equation}
which we call the staggered order in this section. 
In this arrangement, different flavors at the same physical site are put closely with each other, and this is important to control the bond dimension in the computation of DMRG.  
Let us consider another choice for comparison,
\begin{equation}
i=n+N(f-1),
\end{equation}
which we name the flavor order here. 
In this case, we first arrange one of the flavors and then start to arrange the next one, so the flavor degrees of freedom at the same physical sites are separated by $N$, and this clearly violates the above important criterion. 

\begin{figure}[tb]
\centering
\includegraphics[scale=0.45]{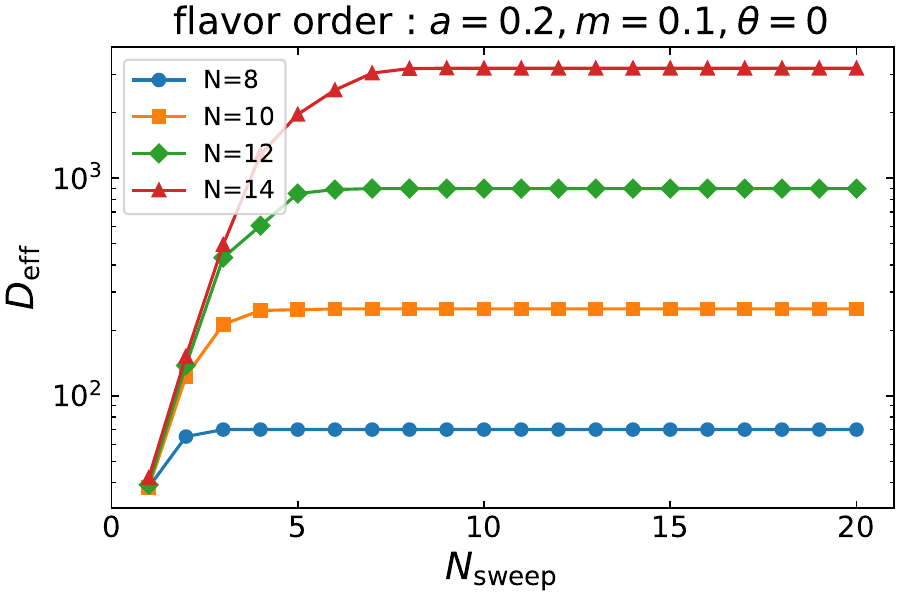}
\includegraphics[scale=0.45]{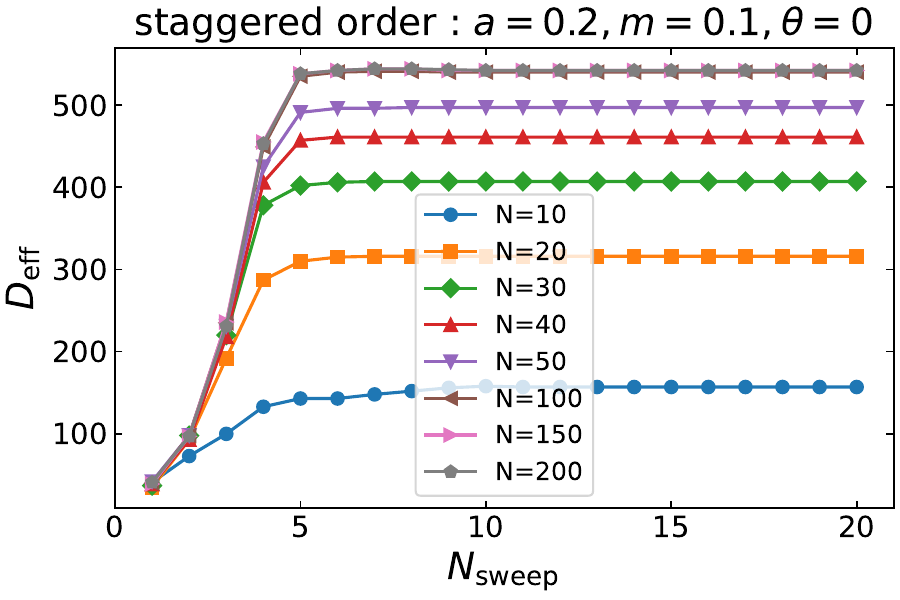}
\caption{\label{fig:Deff_Nswp}
The effective bond dimension $D_{\mathrm{eff}}$
is plotted against the number of sweeps $N_{\mathrm{sweep}}$ for
the flavor order (left) and the staggered order (right). The vertical
axis of the left panel is in log scale, whereas the axis of the right
panel is in linear scale. The lattice spacing and the fermion mass
are set to $a=0.2$ and $m=0.1$.}
\end{figure}

\begin{figure}[tb]
\centering
\includegraphics[scale=0.45]{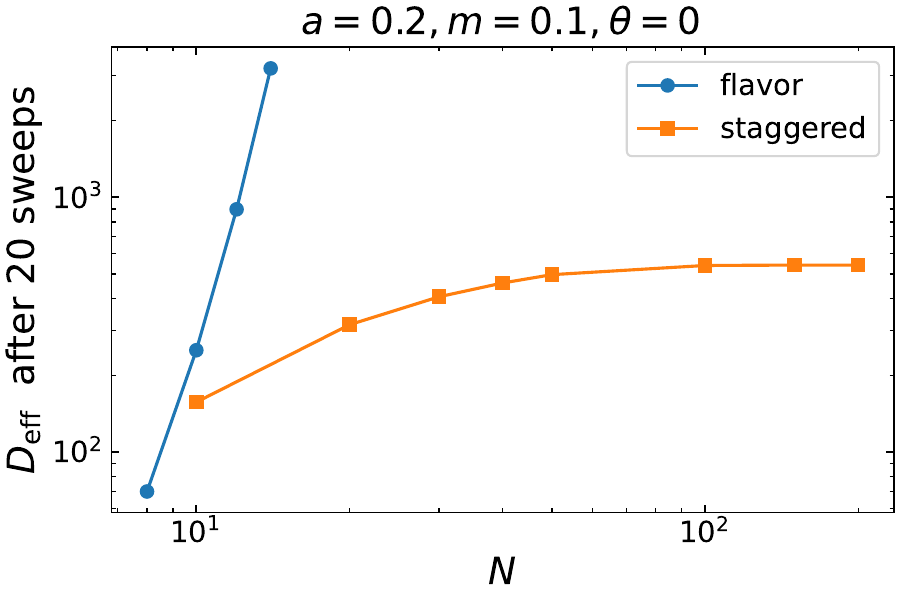}
\caption{\label{fig:Deff_final}
The effective bond dimension $D_{\mathrm{eff}}$ after 20 sweeps is plotted
against the system size $N$ in log-log scale.
The result grows exponentially with $N$ for the flavor order
whereas it is saturated for the staggered order.
The lattice spacing and the fermion mass are set to $a=0.2$ and $m=0.1$.}
\end{figure}

In these two cases, we compare the efficiency of the MPS to represent the ground state
in the gapped phase $\theta=0$.
We obtain the ground state by DMRG and investigate the largest bond dimension
in the MPS, called the effective bond dimension $D_{\mathrm{eff}}$.
The results are plotted against the number of sweeps $N_{\mathrm{sweep}}$
in Fig.~\ref{fig:Deff_Nswp} for various lattice sizes $N$.
We found that $D_{\mathrm{eff}}$ converges around $O(10)$ sweeps for both cases.
However, the dependence on $N$ is totally different.
For the flavor order, the final value of $D_{\mathrm{eff}}$ increases
exponentially with $N$, which is caused by artificial long-range interaction
between the two flavors. On the other hand for the staggered order,
the final value is saturated for sufficiently large $N$.
To show these behaviors, we plot the final values of $D_{\mathrm{eff}}$
after 20 sweeps against $N$ in Fig.~\ref{fig:Deff_final}.

According to Fig.~\ref{fig:Deff_final}, the bond dimension seems to saturate in the case of the staggered order as $N\to \infty$. 
Since the $\ln D_{\mathrm{eff}}$ gives the upper bound for the entanglement entropy, this constant behavior is expected to be the optimal one for the $1+1$d gapped systems. 
On the other hand, $D_{\mathrm{eff}}$ grows exponentially fast for the flavor ordered as $N\to\infty$. 
We suspect that this is because the flavor order puts the entangled flavors in separate locations. 
If we cut the system into two pieces in terms of $i$ with the flavor ordering, the $O(N)$ entangled pairs are cut, and thus the entanglement entropy becomes $O(N)$, which is consistent with the exponential behavior of $D_{\mathrm{eff}}$. 
Therefore, we adopt the staggered order in the whole analysis of this work.

\section{Correlation function in the 1-flavor Schwinger model}
\label{sec:CF_nf1}

We test the validity of the correlation-function scheme in Section~\ref{subsec:result_CF} by examining
the correlation function in the 1-flavor Schwinger model.
When the fermion is massless $m=0$, the model can be analytically solvable and it is equivalent to the free massive boson with mass $\mu'=g/\sqrt{\pi}$.
Thus, this is a good benchmark and we compare the numerical result of DMRG with the analytical answer.

\begin{figure}[tb]
\centering
\includegraphics[scale=0.45]{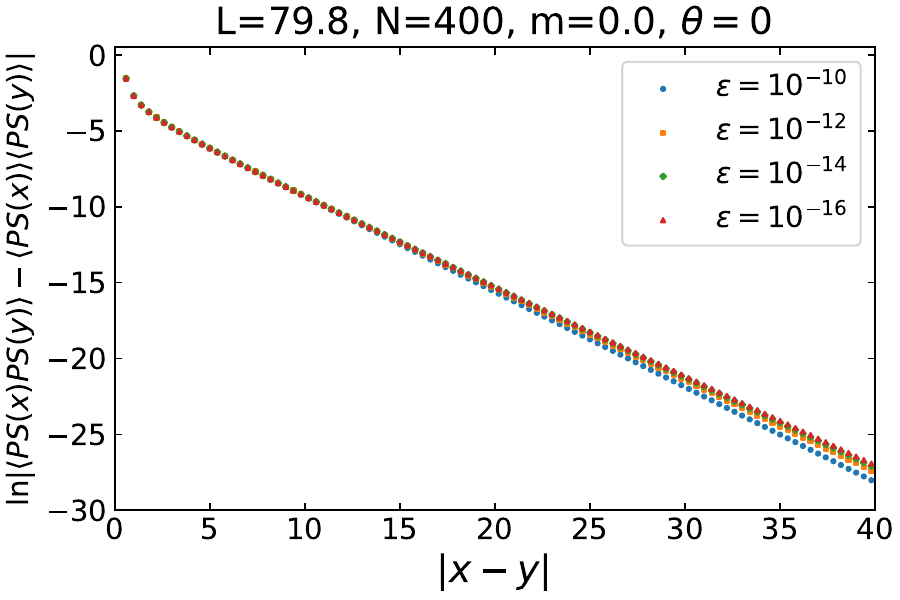}
\includegraphics[scale=0.45]{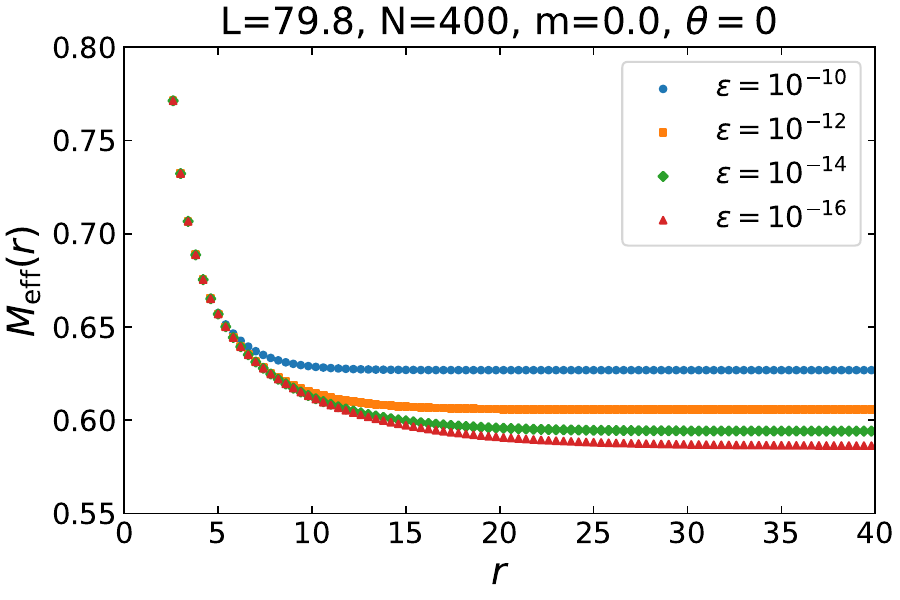}
\caption{\label{fig:cf_nf1}
(Left) The correlation function  $\ln\Braket{PS(x)PS(y)}$ is plotted against the distance $r=|x-y|$
for various values of $\varepsilon$ after subtracting the disconnected part.
The number of lattice sites is $N=400$ and the lattice spacing $a$ is determined so that $L=a(N-1)=79.8$.
(Right) The effective mass $M_{\mathrm{eff}}(r)$ (3-point average) calculated
from the correlation function in the left panel is plotted against $r$.}
\end{figure}

As an analogy of the pseudo scalar meson in the 2-flavor Schwinger model,
we consider the pseudo-scalar operator $PS=-i\bar{\psi}\gamma^{5}\psi$.
The results of the correlation function $\Braket{PS(x)PS(y)}$ are shown
in the left panel of Fig.~\ref{fig:cf_nf1}. Here, the data with different colors
are obtained with the different values of the cutoff parameter $\varepsilon$.
The corresponding effective masses (3-point average) are also plotted
in the right panel of Fig.~\ref{fig:cf_nf1},
where we can see the significant $\varepsilon$ dependence.

\begin{figure}[tb]
\centering
\includegraphics[scale=0.45]{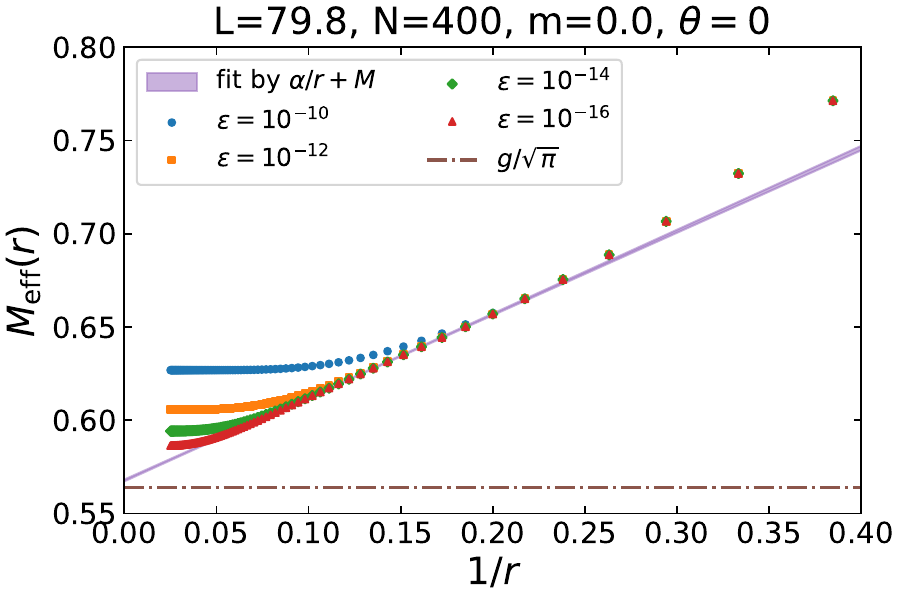}
\caption{\label{fig:fitMeff_nf1}
The effective mass  $M_{\mathrm{eff}}(r)$ is plotted against $1/r$.
The data points for $\varepsilon=10^{-16}$ are fitted by $\alpha/r+M$
inside the region $0.06\protect\leq1/r\protect\leq0.2$.
The fitting result is depicted by the shaded band with systematic error.
The exact mass of the pseudo scalar $g/\sqrt{\pi}$ is also shown by the horizontal broken line.}
\end{figure}

To see the $1/r$ correction of the effective mass, we plot $M_{\mathrm{eff}}(r)$ against $1/r$ in Fig.~\ref{fig:fitMeff_nf1}.
Then we found that the result approaches the expected asymptotic behavior $M_{\mathrm{eff}}(r)\sim\alpha/r+M$
only if the cutoff $\varepsilon$ is sufficiently small.
We fitted the data points for $\varepsilon=10^{-16}$ by $\alpha/r+M$
in the range $0.06\leq1/r\leq0.2$ and obtained $M=0.5677(5)$ and $\alpha=0.446(4)$.
Here the systematic error from the uncertainty of the fitting range is evaluated as explained in Section~\ref{subsec:result_CF}.
We note that this is the result on the finite lattice before taking the continuum limit,
but it turned out to be close to the exact value $M=g/\sqrt{\pi}\approx0.56419$ of the continuum theory.

Therefore, it is quite important to discuss the cutoff (or bond-dimension) dependence especially when we use the correlation-function scheme. 
Indeed, if we naively read the plateau value at $\varepsilon=10^{-10}$, we got an incorrect answer $M\sim0.63$ without observing the $1/\sqrt{r}$ contribution in the Yukawa potential at all.

% \paragraph{Note added.}
% This is also a good position for notes added after the paper has been written.

\bibliographystyle{JHEP}
\bibliography{./Nf2_Schwinger.bib,./QFT.bib}

\end{document}